\documentclass[11pt,a4paper]{article}
\usepackage{jheppub}
\usepackage{comment}
\usepackage{tikz}
\usetikzlibrary{decorations.pathmorphing,decorations.markings}

\tikzset{
  psi/.style={
    decoration={
      markings,
      mark=at position 0.6 with {\arrow{>}}
    },
    postaction={decorate},
    double,
    double distance=1pt
  },
  psiNoArrow/.style={
    decoration={
      markings,
      mark=at position 0.6 with 
    },
    postaction={decorate},
    double,
    double distance=1pt
  },
  nucleon/.style={
    decoration={
      markings,
      mark=at position 0.6 with {\arrow{>}}
    },
    postaction={decorate}
  },
  external/.style={},  %{circle,fill,inner sep=1.5pt}
  gluon/.style={
  decorate, draw=black, 
  decoration={coil,amplitude=4pt, segment length=5pt}
  },
  particle/.style={draw=black, postaction={decorate}, decoration={markings,mark=at position .5 with {\arrow[draw=black]{>}}}},
 photon/.style={decorate, decoration={snake,amplitude=2pt, segment length=5pt}, draw=black}
}

\newcommand{\propagatorOne}{%
  \begin{tikzpicture}[baseline=-3pt]
    \draw[psi] (0, 0) -- (2, 0);
    \draw (0.9, 0.25) -- (1.23, 0.25);
    \draw (1.18, 0.3) -- (1.23, 0.25);
    \draw (1.18, 0.2) -- (1.23, 0.25);
    
    \fill[black] (0,0) circle (0.06cm);
    \fill[black] (2,0) circle (0.06cm);
    
    \node at (0, 0.3) [external]{$(\dot{a})$};
    \node at (2, 0.3) [external]{$(a)$};
    \node at (1, 0.5) [external]{$p$};
  \end{tikzpicture}
}

\newcommand{\propagatorTwo}{%
  \begin{tikzpicture}[baseline=-3pt]
    \draw[psi] (2, 0) -- (0, 0);
    \draw (0.8+0.1, 0.25) -- (1.13+0.1, 0.25);
    \draw (1.08+0.1, 0.3) -- (1.13+0.1, 0.25);
    \draw (1.08+0.1, 0.2) -- (1.13+0.1, 0.25);
    
    \fill[black] (0,0) circle (0.06cm);
    \fill[black] (2,0) circle (0.06cm);
    
    \node at (0, 0.3) [external]{$(a)$};
    \node at (2, 0.3) [external]{$(\dot{a})$};
    \node at (1, 0.5) [external]{$p$};
  \end{tikzpicture}
}

\newcommand{\propagatorThree}{%
  \begin{tikzpicture}[baseline=-3pt]
    \draw[psi] (1, 0) -- (0, 0);
    \draw[psi] (1, 0) -- (2, 0);
    
    \fill[black] (0,0) circle (0.06cm);
    \fill[black] (2,0) circle (0.06cm);
    
    \node at (0, 0.3) [external]{$(b)$};
    \node at (2, 0.3) [external]{$(a)$};
  \end{tikzpicture}
}

\newcommand{\propagatorFour}{%
  \begin{tikzpicture}[baseline=-3pt]
    \draw[psi] (0, 0) -- (1, 0);
    \draw[psi] (2, 0) -- (1, 0);
    
    \fill[black] (0,0) circle (0.06cm);
    \fill[black] (2,0) circle (0.06cm);
    
    \node at (0, 0.3) [external]{$(\dot{b})$};
    \node at (2, 0.3) [external]{$(\dot{a})$};
  \end{tikzpicture}
}

\newcommand{\incomingLeft}{%
  \begin{tikzpicture}[baseline=-3pt]
    \draw[psi] (0, 0) -- (2, 0);
    \draw (0.8+0.1, 0.25) -- (1.13+0.1, 0.25);
    \draw (1.08+0.1, 0.3) -- (1.13+0.1, 0.25);
    \draw (1.08+0.1, 0.2) -- (1.13+0.1, 0.25);
    
    \fill[black] (2,0) circle (0.06cm);
   
    \node at (2, 0.3) [external]{$(a)$};
    \node at (1, 0.5) [external]{$p,\sigma$};
  \end{tikzpicture}
}

\newcommand{\incomingRight}{%
  \begin{tikzpicture}[baseline=-3pt]
    \draw[psi] (2, 0) -- (0, 0);
    \draw (0.8+0.1, 0.25) -- (1.13+0.1, 0.25);
    \draw (1.08+0.1, 0.3) -- (1.13+0.1, 0.25);
    \draw (1.08+0.1, 0.2) -- (1.13+0.1, 0.25);
    
    \fill[black] (2,0) circle (0.06cm);
   
    \node at (2, 0.3) [external]{$(\dot{a})$};
    \node at (1, 0.5) [external]{$p,\sigma$};
  \end{tikzpicture}
}

\newcommand{\outgoingLeft}{%
  \begin{tikzpicture}[baseline=-3pt]
    \draw[psi] (0, 0) -- (2, 0);
    \draw (0.8+0.1, 0.25) -- (1.13+0.1, 0.25);
    \draw (1.08+0.1, 0.3) -- (1.13+0.1, 0.25);
    \draw (1.08+0.1, 0.2) -- (1.13+0.1, 0.25);
    
    \fill[black] (0,0) circle (0.06cm);
   
    \node at (0, 0.3) [external]{$(\dot{a})$};
    \node at (1, 0.5) [external]{$p,\sigma$};
  \end{tikzpicture}
}

\newcommand{\outgoingRight}{%
  \begin{tikzpicture}[baseline=-3pt]
    \draw[psi] (2, 0) -- (0, 0);
    \draw (0.8+0.1, 0.25) -- (1.13+0.1, 0.25);
    \draw (1.08+0.1, 0.3) -- (1.13+0.1, 0.25);
    \draw (1.08+0.1, 0.2) -- (1.13+0.1, 0.25);
    
    \fill[black] (0,0) circle (0.06cm);
   
    \node at (0, 0.3) [external]{$(a)$};
    \node at (1, 0.5) [external]{$p,\sigma$};
  \end{tikzpicture}
}

\newcommand{\psieLeRnuL}{%
  \begin{tikzpicture}[baseline=-3pt]
    \draw[psi] (-0.7, 0.7) -- (0, 0);
    \draw[particle] (-0.7, -0.7) -- (0, 0);
    \draw[particle] (0.7, 0.7) -- (0, 0);
    \draw[particle] (0.7, -0.7) -- (0, 0);
    
    \node at (0.9, 0.9) [external]{$e_{Rf}^{k\ast}$};
    \node at (0.9, -0.9) [external]{$\nu_{Le}^{j}$};
    \node at (-0.9, 0.9) [external]{$\psi^{abc}$};
    \node at (-0.9, -0.9) [external]{$e_{Ld}^{i}$};
  \end{tikzpicture}
}

\newcommand{\psiuRdRdR}{%
  \begin{tikzpicture}[baseline=-3pt]
    \draw[psi] (-0.7, 0.7) -- (0, 0);
    \draw[particle] (-0.7, -0.7) -- (0, 0);
    \draw[particle] (0.7, 0.7) -- (0, 0);
    \draw[particle] (0.7, -0.7) -- (0, 0);
    
    \node at (0.9, 0.9) [external]{$d_{RKf}^{k\ast}$};
    \node at (0.9, -0.9) [external]{$d_{RJe}^{j\ast}$};
    \node at (-0.9, 0.9) [external]{$\psi^{abc}$};
    \node at (-0.9, -0.9) [external]{$u_{LId}^{i\ast}$};
  \end{tikzpicture}
}

\newcommand{\psidLdRnuL}{%
  \begin{tikzpicture}[baseline=-3pt]
    \draw[psi] (-0.7, 0.7) -- (0, 0);
    \draw[particle] (-0.7, -0.7) -- (0, 0);
    \draw[particle] (0.7, 0.7) -- (0, 0);
    \draw[particle] (0.7, -0.7) -- (0, 0);
    
    \node at (0.9, 0.9) [external]{$d_{RJf}^{k\ast}$};
    \node at (0.9, -0.9) [external]{$\nu_{Le}^{j}$};
    \node at (-0.9, 0.9) [external]{$\psi^{abc}$};
    \node at (-0.9, -0.9) [external]{$d_{LId}^{i}$};
  \end{tikzpicture}
}

\newcommand{\psiuLdReL}{%
  \begin{tikzpicture}[baseline=-3pt]
    \draw[psi] (-0.7, 0.7) -- (0, 0);
    \draw[particle] (-0.7, -0.7) -- (0, 0);
    \draw[particle] (0.7, 0.7) -- (0, 0);
    \draw[particle] (0.7, -0.7) -- (0, 0);
    
    \node at (0.9, 0.9) [external]{$d_{RJf}^{k\ast}$};
    \node at (0.9, -0.9) [external]{$e_{Le}^{j}$};
    \node at (-0.9, 0.9) [external]{$\psi^{abc}$};
    \node at (-0.9, -0.9) [external]{$u_{LId}^{i}$};
  \end{tikzpicture}
}

\newcommand{\psihnuL}{%
  \begin{tikzpicture}[baseline=-3pt]
    \draw[psi] (-0.7, 0) -- (0, 0);
    \draw[particle] (0.7, 0.7) -- (0, 0);
    \draw[dashed] (0.7, -0.7) -- (0, 0);
    
    \node at (0.9, 0.9) [external]{$\nu_{Ld}^{i}$};
    \node at (0.9, -0.9) [external]{$h$};
    \node at (-1.1, 0) [external]{$\psi^{abc}$};
    
    \node at (0.3, 0.6) [external]{$q_1$};
    \node at (0.3, -0.6) [external]{$q_2$};
  \end{tikzpicture}
}

\newcommand{\psiZnuL}{%
  \begin{tikzpicture}[baseline=-3pt]
    \draw[psi] (-0.7, 0) -- (0, 0);
    \draw[particle] (0.7, 0.7) -- (0, 0);
    \draw[photon] (0.7, -0.7) -- (0, 0);
    
    \node at (0.9, 0.9) [external]{$\nu_{Ld}^{i}$};
    \node at (0.9, -0.9) [external]{$Z^\nu$};
    \node at (-1.1, 0) [external]{$\psi^{abc}$};
    
    \node at (0.3, 0.6) [external]{$q_1$};
    \node at (0.3, -0.6) [external]{$q_2$};
  \end{tikzpicture}
}

\newcommand{\psiWpluseL}{%
  \begin{tikzpicture}[baseline=-3pt]
    \draw[psi] (-0.7, 0) -- (0, 0);
    \draw[particle] (0.7, 0.7) -- (0, 0);
    \draw[photon] (0.7, -0.7) -- (0, 0);
    
    \node at (0.9, 0.9) [external]{$e_{Ld}^{i}$};
    \node at (0.9, -0.9) [external]{$W^{+\nu}$};
    \node at (-1.1, 0) [external]{$\psi^{abc}$};
    
    \node at (0.3, 0.6) [external]{$q_1$};
    \node at (0.3, -0.6) [external]{$q_2$};
  \end{tikzpicture}
}

\newcommand{\psiAnuL}{%
  \begin{tikzpicture}[baseline=-3pt]
    \draw[psi] (-0.7, 0) -- (0, 0);
    \draw[particle] (0.7, 0.7) -- (0, 0);
    \draw[photon] (0.7, -0.7) -- (0, 0);
    
    \node at (0.9, 0.9) [external]{$\nu_{Ld}^{i}$};
    \node at (0.9, -0.9) [external]{$A^\nu$};
    \node at (-1.1, 0) [external]{$\psi^{abc}$};
    
    \node at (0.3, 0.6) [external]{$q_1$};
    \node at (0.3, -0.6) [external]{$q_2$};
  \end{tikzpicture}
}

\newcommand{\SpinTwopsiZZ}{%
  \begin{tikzpicture}[baseline=-3pt]
    \draw[psi] (-0.7, 0) -- (0, 0);
    \draw[photon] (0.7, 0.7) -- (0, 0);
    \draw[photon] (0.7, -0.7) -- (0, 0);
    
    \node at (0.9, 0.9) [external]{$Z_\nu$};
    \node at (0.9, -0.9) [external]{$Z_\lambda$};
    \node at (-1.1, 0+0.05) [external]{$\psi^{abcd}$};
    
    \node at (0.3, 0.6) [external]{$q_1$};
    \node at (0.3, -0.6) [external]{$q_2$};
  \end{tikzpicture}
}

\newcommand{\SpinTwopsiAZ}{%
  \begin{tikzpicture}[baseline=-3pt]
    \draw[psi] (-0.7, 0) -- (0, 0);
    \draw[photon] (0.7, 0.7) -- (0, 0);
    \draw[photon] (0.7, -0.7) -- (0, 0);
    
    \node at (0.9, 0.9) [external]{$A_\nu$};
    \node at (0.9, -0.9) [external]{$Z_\lambda$};
    \node at (-1.1, 0+0.05) [external]{$\psi^{abcd}$};
    
    \node at (0.3, 0.6) [external]{$q_1$};
    \node at (0.3, -0.6) [external]{$q_2$};
  \end{tikzpicture}
}

\newcommand{\SpinTwopsiAA}{%
  \begin{tikzpicture}[baseline=-3pt]
    \draw[psi] (-0.7, 0) -- (0, 0);
    \draw[photon] (0.7, 0.7) -- (0, 0);
    \draw[photon] (0.7, -0.7) -- (0, 0);
    
    \node at (0.9, 0.9) [external]{$A_\nu$};
    \node at (0.9, -0.9) [external]{$A_\lambda$};
    \node at (-1.1, 0+0.05) [external]{$\psi^{abcd}$};
    
    \node at (0.3, 0.6) [external]{$q_1$};
    \node at (0.3, -0.6) [external]{$q_2$};
  \end{tikzpicture}
}

\newcommand{\SpinTwopsiWplusWminus}{%
  \begin{tikzpicture}[baseline=-3pt]
    \draw[psi] (-0.7, 0) -- (0, 0);
    \draw[photon] (0.7, 0.7) -- (0, 0);
    \draw[photon] (0.7, -0.7) -- (0, 0);
    
    \node at (0.9, 0.9) [external]{$W^+_\nu$};
    \node at (0.9, -0.9) [external]{$W^-_\lambda$};
    \node at (-1.1, 0+0.05) [external]{$\psi^{abcd}$};
    
    \node at (0.3, 0.6) [external]{$q_1$};
    \node at (0.3, -0.6) [external]{$q_2$};
  \end{tikzpicture}
}

\newcommand{\SpinTwopsigg}{%
  \begin{tikzpicture}[baseline=-3pt]
    \draw[psi] (-0.7, 0) -- (0, 0);
    \draw[gluon] (0.7, 0.7) -- (0, 0);
    \draw[gluon] (0.7, -0.7) -- (0, 0);
    
    \node at (0.9, 0.9) [external]{$g_{A\nu}$};
    \node at (0.9, -0.9) [external]{$g_{B\lambda}$};
    \node at (-1.1, 0+0.05) [external]{$\psi^{abcd}$};
    
    \node at (0.15, 0.6) [external]{$q_1$};
    \node at (0.15, -0.6) [external]{$q_2$};
  \end{tikzpicture}
}

\newcommand{\SpinThreepsiAAA}{%
  \begin{tikzpicture}[baseline=-3pt]
    \draw[psi] (-0.7, 0) -- (0, 0);
    \draw[photon] (0.7, 0.7) -- (0, 0);
    \draw[photon] (1.0, 0) -- (0, 0);
    \draw[photon] (0.7, -0.7) -- (0, 0);
    
    \node at (0.9, 0.9) [external]{$A_\nu$};
    \node at (0.9, -0.9) [external]{$A_\epsilon$};
     \node at (1.3, 0) [external]{$A_\sigma$};
    \node at (-1.1, 0+0.08) [external]{$\psi^{abcdef}$};
    
    \node at (0.3, 0.6) [external]{$q_1$};
    \node at (0.9, -0.25) [external]{$q_2$};
    \node at (0.3, -0.6) [external]{$q_3$};
  \end{tikzpicture}
}

\newcommand{\SpinThreepsiZZZ}{%
  \begin{tikzpicture}[baseline=-3pt]
    \draw[psi] (-0.7, 0) -- (0, 0);
    \draw[photon] (0.7, 0.7) -- (0, 0);
    \draw[photon] (1.0, 0) -- (0, 0);
    \draw[photon] (0.7, -0.7) -- (0, 0);
    
    \node at (0.9, 0.9) [external]{$Z_\nu$};
    \node at (0.9, -0.9) [external]{$Z_\epsilon$};
     \node at (1.3, 0) [external]{$Z_\sigma$};
    \node at (-1.1, 0+0.08) [external]{$\psi^{abcdef}$};
    
    \node at (0.3, 0.6) [external]{$q_1$};
    \node at (0.9, -0.25) [external]{$q_2$};
    \node at (0.3, -0.6) [external]{$q_3$};
  \end{tikzpicture}
}

\newcommand{\SpinThreepsiAAZ}{%
  \begin{tikzpicture}[baseline=-3pt]
    \draw[psi] (-0.7, 0) -- (0, 0);
    \draw[photon] (0.7, 0.7) -- (0, 0);
    \draw[photon] (1.0, 0) -- (0, 0);
    \draw[photon] (0.7, -0.7) -- (0, 0);
    
    \node at (0.9, 0.9) [external]{$A_\nu$};
    \node at (1.3, 0) [external]{$A_\sigma$};
    \node at (0.9, -0.9) [external]{$Z_\epsilon$};
    \node at (-1.1, 0+0.08) [external]{$\psi^{abcdef}$};
    
    \node at (0.3, 0.6) [external]{$q_1$};
    \node at (0.9, -0.25) [external]{$q_2$};
    \node at (0.3, -0.6) [external]{$q_3$};
  \end{tikzpicture}
}

\newcommand{\SpinThreepsiAZZ}{%
  \begin{tikzpicture}[baseline=-3pt]
    \draw[psi] (-0.7, 0) -- (0, 0);
    \draw[photon] (0.7, 0.7) -- (0, 0);
    \draw[photon] (1.0, 0) -- (0, 0);
    \draw[photon] (0.7, -0.7) -- (0, 0);
    
    \node at (0.9, 0.9) [external]{$A_\nu$};
    \node at (1.3, 0) [external]{$Z_\sigma$};
    \node at (0.9, -0.9) [external]{$Z_\epsilon$};
    \node at (-1.1, 0+0.08) [external]{$\psi^{abcdef}$};
    
    \node at (0.3, 0.6) [external]{$q_1$};
    \node at (0.9, -0.25) [external]{$q_2$};
    \node at (0.3, -0.6) [external]{$q_3$};
  \end{tikzpicture}
}

\newcommand{\SpinThreepsiAWW}{%
  \begin{tikzpicture}[baseline=-3pt]
    \draw[psi] (-0.7, 0) -- (0, 0);
    \draw[photon] (0.7, 0.7) -- (0, 0);
    \draw[photon] (1.0, 0) -- (0, 0);
    \draw[photon] (0.7, -0.7) -- (0, 0);
    
    \node at (0.9, 0.9) [external]{$A_\nu$};
    \node at (1.3, 0) [external]{$W^+_\sigma$};
    \node at (0.9, -0.9) [external]{$W^-_\epsilon$};
    \node at (-1.1, 0+0.08) [external]{$\psi^{abcdef}$};
    
    \node at (0.3, 0.6) [external]{$q_1$};
    \node at (0.9, -0.25) [external]{$q_2$};
    \node at (0.3, -0.6) [external]{$q_3$};
  \end{tikzpicture}
}

\newcommand{\SpinThreepsiZWW}{%
  \begin{tikzpicture}[baseline=-3pt]
    \draw[psi] (-0.7, 0) -- (0, 0);
    \draw[photon] (0.7, 0.7) -- (0, 0);
    \draw[photon] (1.0, 0) -- (0, 0);
    \draw[photon] (0.7, -0.7) -- (0, 0);
    
    \node at (0.9, 0.9) [external]{$Z_\nu$};
    \node at (1.3, 0) [external]{$W^+_\sigma$};
    \node at (0.9, -0.9) [external]{$W^-_\epsilon$};
    \node at (-1.1, 0+0.08) [external]{$\psi^{abcdef}$};
    
    \node at (0.3, 0.6) [external]{$q_1$};
    \node at (0.9, -0.25) [external]{$q_2$};
    \node at (0.3, -0.6) [external]{$q_3$};
  \end{tikzpicture}
}

\newcommand{\SpinThreepsiAGG}{%
  \begin{tikzpicture}[baseline=-3pt]
    \draw[psi] (-0.7, 0) -- (0, 0);
    \draw[photon] (0.7, 0.7) -- (0, 0);
    \draw[gluon] (1.0, 0) -- (0, 0);
    \draw[gluon] (0.7, -0.7) -- (0, 0);
    
    \node at (0.9, 0.9) [external]{$A_\nu$};
    \node at (1.3, 0) [external]{$G_\sigma$};
    \node at (0.9, -0.9) [external]{$G_\epsilon$};
    \node at (-1.1, 0+0.08) [external]{$\psi^{abcdef}$};
    
    \node at (0.27, 0.7) [external]{$q_1$};
    \node at (0.9, -0.35) [external]{$q_2$};
    \node at (0.25, -0.7) [external]{$q_3$};
  \end{tikzpicture}
}

\newcommand{\SpinThreepsiZGG}{%
  \begin{tikzpicture}[baseline=-3pt]
    \draw[psi] (-0.7, 0) -- (0, 0);
    \draw[photon] (0.7, 0.7) -- (0, 0);
    \draw[gluon] (1.0, 0) -- (0, 0);
    \draw[gluon] (0.7, -0.7) -- (0, 0);
    
    \node at (0.9, 0.9) [external]{$Z_\nu$};
    \node at (1.3, 0) [external]{$G_\sigma$};
    \node at (0.9, -0.9) [external]{$G_\epsilon$};
    \node at (-1.1, 0+0.08) [external]{$\psi^{abcdef}$};
    
    \node at (0.27, 0.7) [external]{$q_1$};
    \node at (0.9, -0.35) [external]{$q_2$};
    \node at (0.25, -0.7) [external]{$q_3$};
  \end{tikzpicture}
}

\newcommand{\SpinThreepsiGGG}{%
  \begin{tikzpicture}[baseline=-3pt]
    \draw[psi] (-0.7, 0) -- (0, 0);
    \draw[gluon] (0.7, 0.7) -- (0, 0);
    \draw[gluon] (1.0, 0) -- (0, 0);
    \draw[gluon] (0.7, -0.7) -- (0, 0);
    
    \node at (0.9, 0.9) [external]{$G_\nu$};
    \node at (1.3, 0) [external]{$G_\sigma$};
    \node at (0.9, -0.9) [external]{$G_\epsilon$};
    \node at (-1.1, 0+0.08) [external]{$\psi^{abcdef}$};
    
    \node at (0.27, 0.7) [external]{$q_1$};
    \node at (0.9, -0.35) [external]{$q_2$};
    \node at (0.25, -0.7) [external]{$q_3$};
  \end{tikzpicture}
}

\newcommand{\psiTohnuExample}{%
  \begin{tikzpicture}[baseline=-3pt]
    \draw[psi] (-0.7, 0) -- (0, 0);
    \draw[particle] (0.7, 0.7) -- (0, 0);
    \draw[dashed] (0.7, -0.7) -- (0, 0);
    
    \node at (0.9, 1.0) [external]{$\nu_{Ld}(q_1)$};
    \node at (0.9, -0.9) [external]{$h(q_2)$};
    \node at (-1.35, 0) [external]{$\psi^{abc}(p)$};
    
  \end{tikzpicture}
}

\newcommand{\psiTohbarnuExample}{%
  \begin{tikzpicture}[baseline=-3pt]
    \draw[psi] (0, 0) -- (-0.7, 0);
    \draw[particle] (0, 0) -- (0.7, 0.7);
    \draw[dashed] (0.7, -0.7) -- (0, 0);
    
    \node at (0.9, 1.0) [external]{$\nu_{L\dot{d}}^{\ast}(q_1)$};
    \node at (0.9, -0.9) [external]{$h(q_2)$};
    \node at (-1.35, 0) [external]{$\psi^{\dot{a}\dot{b}\dot{c}}(p)$};
    
  \end{tikzpicture}
}

\usepackage{hyperref}
\usepackage{amsmath,amsfonts,amssymb}
\usepackage{booktabs}
\usepackage{enumitem}
\usepackage{graphicx}
\usepackage{siunitx}
\usepackage{xcolor}
\usepackage{verbatim}
\setlist[itemize]{leftmargin=*}

\newcommand{\be}{\begin{equation}}
\newcommand{\ee}{\end{equation}}
\newcommand{\bea}{\begin{equation}\begin{aligned}}
\newcommand{\eea}{\end{aligned}\end{equation}}
\newcommand{\td}{{\rm d}}

\newcommand\Mcal{\mathcal{M}}

\newcommand{\MeV}{\rm{MeV}}
\newcommand{\GeV}{\rm{GeV}}
\newcommand{\TeV}{\rm{TeV}}

\newcommand{\sW}{\sin\theta_W}
\newcommand{\cW}{\cos\theta_W}

\newcommand{\hc}{\text{h.c.}}
\newcommand{\ie}{{\it i.e.}}
\newcommand{\eg}{{\it e.g.}}

\newcommand{\gsim}{\lower.7ex\hbox{$\;\stackrel{\textstyle>}{\sim}\;$}}
\newcommand{\lsim}{\lower.7ex\hbox{$\;\stackrel{\textstyle<}{\sim}\;$}}

\definecolor{grey}{cmyk}{0,0,0,0.75}
\definecolor{tangerine}{cmyk}{0,0.5,1,0}
\definecolor{darkgreen}{cmyk}{1,0,1,0.23}
\definecolor{Red}{rgb}{1,0,0}
\definecolor{Blue}{rgb}{0,0,1}
\definecolor{Green}{rgb}{0,1,0}
\definecolor{Grey}{cmyk}{0,0,0,0.75}
\definecolor{Tangerine}{cmyk}{0,0.5,1,0}
\definecolor{Darkgreen}{cmyk}{1,0,1,0.23}
\definecolor{Cyan}{cmyk}{1,0,0,0}
\definecolor{Yellow}{cmyk}{0,0,1,0}
\definecolor{darkblue}{cmyk}{1,0.69,0,0.11}

\newcommand{\orange}[1]{\textcolor{Tangerine}{#1}}

\newcommand{\nicpb}{Laboratory of High Energy and Computational Physics, NICPB, R\"avala pst. 10, 10143 Tallinn, Estonia}
\newcommand{\ippp}{Institute for Particle Physics Phenomenology, Department of Physics, Durham University, Durham DH1 3LE, United Kingdom}
\newcommand{\capfe}{CAFPE and Departamento de F\'isica Te\'orica y del Cosmos, Universidad de Granada, E-18071 Granada, Spain}
\newcommand{\lapth}{Universit\'e Savoie--Mont Blanc, USMB, CNRS, LAPTh, F-74000 Annecy, France}
\newcommand{\epem}{e^+e^-}

\begin{document}

\title{Higher-spin particles at high-energy colliders}

\author[a]{Juan C. Criado,}
\author[b,c,d]{Abdelhak Djouadi,}
\author[b]{Niko Koivunen,}
\author[b]{Martti Raidal,}
\author[b]{and Hardi~Veerm\"{a}e}  
\affiliation[a]{\ippp}  
\affiliation[b]{\nicpb} 
\affiliation[c]{\capfe}  
\affiliation[d]{\lapth}

\emailAdd{juan.c.criado@durham.ac.uk}
\emailAdd{abdelhak.djouadi@cern.ch}
\emailAdd{niko.koivunen@kbfi.ee} 
\emailAdd{martti.raidal@cern.ch}
\emailAdd{hardi.veermae@cern.ch}

\abstract{
Using an effective field theory approach for higher-spin fields, we derive the interactions of colour singlet and electrically neutral particles with a spin higher than unity, concentrating on the spin-3/2, spin-2, spin-5/2 and spin-3 cases. We compute the decay rates and production cross sections in the main channels for spin-3/2 and spin-2 states at both electron-positron and hadron colliders, and identify the most promising novel experimental signatures for discovering such particles at the LHC. The discussion is qualitatively extended to the spin-5/2 and spin-3 cases. Higher-spin particles exhibit a rich phenomenology and have signatures that often resemble the ones of supersymmetric and extra-dimensional theories. To enable further studies of higher-spin particles at collider and beyond, we collect the relevant Feynman rules and other technical details.
}

\maketitle

%%%%%%%%%%%%%%%%%%%%%%%%%%%%%%%%%%%%%%%%%%%%%%%%%%%%%%%%%%%%%%%%%%%%%%%%%%%%%
\section{Introduction}
%%%%%%%%%%%%%%%%%%%%%%%%%%%%%%%%%%%%%%%%%%%%%%%%%%%%%%%%%%%%%%%%%%%%%%%%%%%%%

Testing the standard model (SM) of particle physics and searching for new phenomena beyond it is the main objective of high-energy colliders such as the Large Hadron Collider (LHC) at CERN. For decades, this search has been focused on new particles predicted by theories that address the hierarchy problem of the SM and naturally explain the unbearable lightness of the Higgs boson \cite{Aad:2012tfa,Chatrchyan:2012ufa}. Among these, supersymmetric theories~\cite{Wess:1974tw,Fayet:1977vd} and models with extra space-time dimensions~\cite{Antoniadis:1998ig,Randall:1999ee} have played a more than considerable role. The vast majority of these beyond the SM scenarios predict new particles that are similar to the existing ones and, up to few notable exceptions related to gravity, have a spin smaller than or equal to unity: new scalar (often Higgs) bosons, additional spin-$1/2$ quarks and leptons and extra vector (mainly gauge) bosons. This is the case for supersymmetric theories, which, for instance, predict spin-0 partners for the SM fermions and spin-$1/2$ ones for the known gauge bosons as well as for the Higgs bosons, and for extra space-time dimensional models which predict a tower of same-spin Kaluza-Klein excitations for all the SM particles.

Now that the LHC has set stringent lower bounds on the masses of the above two sets of new particles, well above the TeV scale for most of them, to the extent where the two theories appear to be less ``natural", one may adopt (or return to) an agnostic attitude towards new physics. In particular, one should stay as model-independent as possible and search for new particles of any type, considering more seriously phenomena that originate from scenarios not related to the solution of the hierarchy problem and, hence, according to old standards, have less theoretical motivation. Such models could lead to a rather rich phenomenology, predict particles with more exotic quantum numbers, and suggest novel signatures that might not be yet searched for by experimental collaborations. Especially, there may exist particles with a spin higher than unity. In fact, candidates for such states have been predicted by the two leading theories mentioned above. 

Indeed, a well-known example of a spin-3/2 fermion is the gravitino, the graviton's supersymmetric partner, that naturally appears in supergravity~\cite{Fayet:1977vd} and gauge mediated~\cite{Dine:1994vc} supersymmetry breaking models. Such scenarios were explored at length in the last decades, in particular as these new particles were considered to be good dark matter (DM) candidates. However, in both scenarios, only the spin-1/2 (or longitudinal) component of the gravitino couples sufficiently strongly to ordinary matter so that, in practice and at least for collider physics purposes, the gravitino behaves essentially as a spin-1/2 fermion\footnote{In fact, a longitudinal gravitino is simply the Goldstino that signals the spontaneous breaking of global supersymmetry~\cite{Fayet:1977vd,Casalbuoni:1988kv}, whose coupling is inversely proportional to the supersymmetry breaking scale, given by the square-root of the gravitino mass times the Planck mass. In general, this leads to very light gravitinos, which we will not consider here.}. 

Heavy spin-2 particles have been discussed in the context of extra space-time dimensional scenarios~\cite{Antoniadis:1998ig,Randall:1999ee}. They appear as the massive Kaluza-Klein excitations of the massless graviton. However, the interactions of these heavy gravitons are very specific as, for instance, they couple universally to all particles~\cite{Giudice:1998ck,Han:1998sg}. Massive spin-2 particles also appear in extensions of General Relativity, \eg, in theories of bi-metric gravity~\cite{Hassan:2011zd}. However, these particles also have universal, gravity-strength interactions. These states may thus play the role of DM~\cite{Babichev:2016hir,Babichev:2016bxi,Marzola:2017lbt}, but their interactions are irrelevant for collider physics.

Particles with an even higher spin have been put forward only in very few occurrences. For example, spin-$5/2$ states have been considered in hadronic physics~\cite{Shklyar:2009cx}, while spin-3 particles have been discussed only at the formal Lagrangian level in specific theories as well as in hadronic physics; see, \eg, Refs.~\cite{Bergshoeff:2016soe,jafarzade2021phenomenology} for recent accounts.  

A reason for the lack of studies of generic higher-spin particles arises from problems associated with their nature: the absence of a physically meaningful and mathematically consistent framework for performing computations with interacting higher-spin degrees of freedom, as well as the absence of a consistent ultraviolet completion. For instance, most of the work on spin-3/2 particles is conducted in the Rarita-Schwinger framework~\cite{Rarita:1941mf} in which the spin-3/2 field is described as a vector spinor with more components than a physical spin-3/2 particle. Due to certain local symmetries, the unphysical degrees of freedom are eliminated by constraints built into the free Lagrangian. The couplings of the Rarita-Schwinger field must respect these symmetries. Otherwise, the counting of physical degrees of freedom will be inconsistent. 

This fact has been often ignored in nuclear physics when computing the pion-nucleon interaction mediated by a spin-3/2 $\Delta$ resonance~\cite{Peccei:1969zq, Davidson:1991xz, Scholten:1996mw, Korchin:1998ff, Mota:1999qp, Lahiff:1999ur}, for instance, and in collider studies of generic spin-3/2 particles. There have been attempts to cure the problem by rewriting the interactions of the spin-3/2 fields~\cite{Pascalutsa:1998pw, Pascalutsa:1999zz, Pascalutsa:2000kd,Hagen:1982ez,Haberzettl:1998rw,Deser:2000dz, Pilling:2004wk,Pilling:2004cu,Wies:2006rv,Napsuciale:2006wr,Delgado-Acosta:2013kva,DelgadoAcosta:2015ypa,Mart:2019jtb}. Difficulties in formulating $\Delta$-resonance interactions in a Lagrangian description have been discussed in Ref.~\cite{Benmerrouche:1989uc}. In general, Rarita-Schwinger particles with minimal gauge interaction with photons, massive vector bosons and gluons, run into inconsistencies~\cite{Johnson:1960vt} such as causality violation as well as uncontrollable unitarity violating processes at energies not far from the mass scale of the new particles. Most of the studies performed in the past were affected by such problems. Essentially, only supersymmetric theories with specifically fixed couplings and masses have been known to avoid those problems, suggesting that physical spin-3/2 particles should be identified with the gravitino~\cite{Grisaru:1976vm,Grisaru:1977kk}.

Recently, an effective field theory of a generic massive particle of any spin has been developed~\cite{Criado:2020jkp}, following an idea originally proposed by Weinberg~\cite{Weinberg:1964cn}. Although it does not admit a Lagrangian description, this effective theory contains only physical higher-spin degrees of freedom and allows for a consistent computation of physical observables for general-spin particles. It avoids the inconsistencies that often appear in other field-theoretical descriptions of higher spin and reproduces the existing results for low-spin states. As an illustration, this method has been applied successfully to study higher-spin DM particles in terms of a general-spin singlet with symmetric couplings to the Higgs bosons~\cite{Criado:2020jkp}, a setup which automatically arises for higher-spin states\footnote{Generic higher-spin DM states have also been studied recently in Ref.~\cite{Falkowski:2020fsu} using on-shell amplitude methods for massive particles~\cite{Arkani-Hamed:2017jhn}. The approach taken in this paper is different from the latter as relies on Feynman rules allowing to construct scattering amplitudes at an arbitrary level in perturbation theory. Non-relativistic scattering of generic higher-spin DM has recently been considered in Refs.~\cite{Gondolo:2020wge,Gondolo:2021fqo}. Super-heavy higher-spin DM has been studied in Ref.~\cite{Alexander:2020gmv}. The $(j, 0) \oplus (0, j)$ representation of the Lorentz group employed in this work and Ref.~\cite{Criado:2020jkp} has been used in Refs.~\cite{Hernandez-Arellano:2018sen,Hernandez-Arellano:2019qgd} to study spin-1 DM. General theoretical limitations on massive higher spin fields based on the analyticity of the S-matrix have been discussed in Refs.~\cite{Caron-Huot:2016icg,Afkhami-Jeddi:2018apj,Bellazzini:2019bzh}.}.

In this paper, we use the framework of Ref.~\cite{Criado:2020jkp} to study the collider phenomenology of higher-spin particles. In detail, we study higher-spin particles that are singlets under the SM gauge interactions and consider their simplest linear interaction Hamiltonians with SM quarks, leptons and gauge bosons. In this setup, the higher-spin particles are Majorana.  Throughout the paper, we will use effective operators to describe the interactions of the higher-spin fields, that is, $j>1$, that we will denote by $\psi_j$. The lowest order operators linear in $\psi_j$ have dimension $1+3  j $ for bosons and $5/2+3  j $ for fermions, while the lowest order operator quadratic in $\psi_j$ is $\psi^{(a)}\psi_{(a)} |\phi^2| + \hc$, where $\phi$ is the SM Higgs doublet, and has dimension $4 + 2j$. The quadratic operators are thus dominant for spins $j \geq 4$ in the case of bosons and for spins $j \geq 5/2$ in the case of fermions~\cite{Criado:2020jkp}. The higher the spin $j$, the higher the mass dimension of lowest order operators grows. This will generically lead to a steeper energy dependence in the UV and to strongly suppressed interactions below the EFT scale as $j$ increases.

We focus on higher-spin scenarios in which the $\psi$-linear interactions naturally dominate, that is, on spin $3/2$, $2$ and $3$. We will also present some results for spin-$5/2$ as a nontrivial example of an exotic higher-spin fermion. To this aim, we present the effective Hamiltonians describing interactions of $\psi_j$ with the SM fields for all the cases under consideration, $j=3/2,$ $2$, $5/2$ and $3$. However, their collider phenomenology is worked out at different levels of detail depending on the spin. Most of the effort is directed toward studying spin-3/2 and spin-2 particles, for which we compute the $\psi_{3/2}$ and $\psi_{2}$ decay rates and production cross sections both at hadron and lepton colliders. After that, we discuss the most striking experimental signatures and compare those with the ones of supersymmetric and extra-dimensional theories. We shall also discuss the existing constraints on those particles and outline potential future research directions. We extend this discussion to the higher spin-5/2 and spin-3 cases. One of the objectives of this exploratory work is to open the possibility of more extensive studies of higher-spin physics. In order to facilitate this aim, we collect the required technical details in the appendix.

The rest of the paper is organized as follows. In Section~\ref{sec:spin-3/2} we focus on the phenomenology of spin-3/2 particles and, in Section~\ref{sec:spin-2}, on the one of spin-2 particles. After deriving the effective interactions of the particles, we discuss their decay modes and production cross sections and summarise their main signatures at the LHC. The interactions and experimental signatures for spin-5/2 fermions and spin-3 bosons are presented in Section~\ref{sec:spin-5/2} and Section~\ref{sec:spin-3}, respectively. Finally, in Section~\ref{sec:conc}, we discuss other implications of these higher-spin particles, outline future research directions and present our conclusions. The formalism enabling consistent studies of higher-spin particles is outlined in Appendix~\ref{app:formalism}: we list the relevant Feynman rules, discuss the narrow width approximation, and give a detailed example of a computation of higher-spin processes. Throughout the paper we use natural units $\hbar = c = 1$ and the metric signature $(+,\!-,\!-,\!-)$.

%%%%%%%%%%%%%%%%%%%%%%%%%%%%%%%%%%%%%%%%%%%%%%%%%%%%%%%%%%%%%%%%%%%%%%%%%%%%%
\section{Spin-3/2 particles}
\label{sec:spin-3/2}
%%%%%%%%%%%%%%%%%%%%%%%%%%%%%%%%%%%%%%%%%%%%%%%%%%%%%%%%%%%%%%%%%%%%%%%%%%%%%

\subsection{Formalism and interactions}

In high-energy particle physics, both spin-3/2 leptons~\cite{LeiteLopes:1980mh, LeiteLopes:1981ys, Burges:1983zg, Spehler:1986aq, Almeida:1992np, Montero:1993np, Almeida:1995yp,Cakir:2007wn,Eboli:1995uv, Abdullah:2016jji} and spin-3/2 quarks \cite{LeiteLopes:1980pa, Moussallam:1989nm, Dicus:1998yc, Dicus:2012uh, Christensen:2013aua} have been considered in the past. For instance, the production of spin-3/2 particles at hadron colliders has been discussed in Refs.~\cite{Moussallam:1989nm, Dicus:1998yc, Dicus:2012uh,Christensen:2013aua}, while production at lepton colliders has been considered in Refs.~\cite{Almeida:1995yp,Cakir:2007wn}. Indirect effects of spin-3/2 particles, through their virtual exchange in high energy processes, has also been discussed with some examples being the $t$-channel exchange of a spin-3/2 lepton in processes such as $\epem \to 2\gamma$ and $e \gamma \to e\gamma$~\cite{Walsh:1999pb}, and the exchange of spin-3/2 quark in top pair production at hadron colliders, $gg,q\bar q \to t \bar t$~\cite{Stirling:2011ya}. However, it is assumed in most of these analyses that the spin-3/2 particles have colour or electric charge, allowing their pair production in proton-proton or electron-positron collisions. This will not be the case here as such interactions would lead to an unmanageable violation of causality and unitarity~\cite{Johnson:1960vt}.

For example, at hadron colliders, spin-3/2 quarks that couple to gluons could be pair produced in gluon fusion and quark-antiquark annihilation, $gg, q\bar q \to \psi_{3/2} \bar \psi_{3/2}$. The partonic cross sections, which depend only on the known gauge coupling constant $\alpha_s$ and the particle mass $m_{3/2}$, grows with the third power of the partonic centre-of-mass energy, $\hat \sigma \propto \hat s^3$~\cite{Moussallam:1989nm, Dicus:1998yc, Dicus:2012uh,Hassanain:2009at}. Such a steep rise leads to unitarity violation at tree-level for energies of the order of $\approx 7m$~\cite{Hassanain:2009at}. The interaction needs, therefore, to be damped by some form-factors in order to remain viable at these energies. 

To overcome these problems, we consider a generic Majorana spin-3/2 field $\psi_{3/2}$ that is a SM singlet. Therefore, the issues related to gauging the higher-spin fields do not appear here, while the interactions of $\psi_{3/2}$ to gauge bosons are still included. We use the effective field theory approach of Ref.~\cite{Criado:2020jkp}, so we avoid problems related to the existence of unphysical degrees of freedom that appear in other formulations of higher spin. A brief overview of the multispinor formalism can be found in Appendix~\ref{app:formalism}. Since a spin $j$ field carries an effective dimension of $\Delta_\psi = j + 1$, the lowest dimension of the operators linear in a SM-singlet spin-3/2 field is 7. Operators of dimension 5 are allowed when the field has some non-vanishing SM charge. They are always of the form $\psi_j F f$, where $F$ is a SM field-strength tensor, and $f$ is a SM spin-$1/2$ fermion. For example, when $\psi_{3/2}$ is a colour triplet or sextet, $\psi_{3/2} g q$ contact interactions are possible at this level.

For a singlet spin-3/2 field $\psi_{3/2}$, there are 6 independent dimension-7 operators,
\bea\label{eq:L-linear}
	-\mathcal{H}_{\text{linear}} =	\frac{1}{\Lambda^3} \psi_{3/2}^{abc} \Big[ \
&   c^{ijk}_q\epsilon^{IJK} u^{i\ast}_{RIa} d^{j\ast}_{RJb}d^{k\ast}_{RKc}  
+   c^{ijk}_l (L_{La}^{iT}\epsilon L^j_{Lb}) e^{k*}_{Rc}
    \\
&+   c^{ijk}_{lq} (Q_{LIa}^{iT}\epsilon L_{Lb}^j)d_{Rc}^{kI\ast} 
%&	c^\phi (D_{a\dot{a}} \tilde{\phi})^\dagger  D_b{}^{\dot{a}} l_c  
+	c_i^\phi \sigma^{\mu\nu}_{ab} (D_{\mu} \tilde{\phi})^\dagger  D_{\nu} L_{Lc}^i  
    \\
&+	c_i^B \tilde{\phi}^\dagger \sigma^{\mu\nu}_{ab} B_{\mu\nu} L_{Lc}^i
+	c_i^W \tilde{\phi}^\dagger \sigma^{\mu\nu}_{ab} \sigma_n W^n_{\mu\nu} L_{Lc}^i 
\ \Big] 
   + \text{h.c.},
\eea
where $a,b,c$ are two-spinor indices, $i,j,k$ are flavour indices, $I$ and $J$ the colour indices and $n$ is the SU(2)-triplet index. 
We have used integration by parts to eliminate redundant operators, but we have not performed field redefinitions.
The coefficient $c_l^{ijk}$ is symmetric in $ij$, while $c_q^{ijk}$ is symmetric in $jk$. $L^i_a$ and $Q^i_a$ are the left-handed lepton and quark doublets
\be
    L^i_{La}\equiv
\left(\begin{array}{c}
    \nu_{La}^i\\
    e_{La}^i
\end{array}\right),\quad
    Q^i_{LIa}\equiv
\left(\begin{array}{c}
    u_{LIa}^i\\
    d_{LIa}^i
\end{array}\right),
\ee
while $e_R^i$, $u_R^i$ and $d_R^i$ are the right-handed lepton and quark singlets. $B_{\mu \nu}$ and $W_{\mu \nu}$ denote the ${\rm U(1)_Y}$ and ${\rm SU(2)_L}$ field strengths and $\phi$ is the SM Higgs doublet. In the unitary gauge, one has
\be
    \phi=\frac{1}{\sqrt{2}}
    \left(\begin{array}{c}
    0\\
    H+v
    \end{array}\right),
\ee
where $v$ is the vacuum expectation value $v=246$ GeV and $H$ the physical Higgs boson produced at the LHC \cite{Aad:2012tfa,Chatrchyan:2012ufa}. We define $D_{a\dot{a}} = \sigma^\mu_{a\dot{a}} D_\mu$, with $D_\mu$ being the usual 4-vector covariant derivative, and in terms of the identity matrix $\sigma^0$ and Pauli $\sigma^{1,2,3}$ matrices ${(\sigma^{\mu\nu})_a}^b \equiv \frac{i}{4} \left[\sigma^\mu_{a\dot{b}} (\bar{\sigma}^\nu)^{\dot{b}b} - \sigma^\nu_{a\dot{b}} (\bar{\sigma}^\mu)^{\dot{b}b}\right]$.

The Feynman rules for the various interactions are listed in Appendix~\ref{app:feyn}, where we have restricted ourselves to those operators that lead to the dominant processes, namely, to the quartic point-like interaction of the spin-3/2 particle with three fermions and to the triple vertices involving the spin-3/2 particle, a charged lepton or a neutrino and a gauge or Higgs boson; these vertices will give the dominant effects which will be discussed in this paper. One can add a Higgs or a gauge boson line to turn the three-particle vertices involving bosons into four-particle vertices, but, in this case, this interaction will be suppressed by a power of the vacuum expectation value $v$ or by an additional weak gauge coupling. Hence, it will lead to subleading processes, which we will not consider here. 

Another simplification that we will adopt in the following is the absence of flavour violation and, hence, the spin-3/2 field will couple only to fermions of the same generation. For simplicity, we will assume that the lightest new spin-3/2 particle is the one that couples to the first generation quarks and leptons, for which one can safely neglect the masses and the mixing. We restrict our analysis to this case. We further assume that there are no new sources of CP-violation, implying that all couplings of the spin-3/2 field are real. 

In summary, the general Hamiltonian of eq.~\eqref{eq:L-linear} will take a much simpler form in terms of the lepton and quark doublets of the first generation $l^T=(\nu, e)_L$ and $q^T=(u,d)_L$,
\bea
\label{eq:L-simple}
	-\mathcal{H}_{\text{linear}} 	=
	\frac{1}{\Lambda^3} \psi^{abc}_{3/2} \Big[ \
&   c_q \epsilon^{IJK} u^*_{Ia} d^*_{Jb} d^*_{Kc} 	
+   c_l (l^T_a \epsilon l_b) e^*_c
+ 	c_{lq} (q^T_{Ia}\epsilon l_b) d^{*I}_c  \\
+&	c_B \tilde{\phi}^\dagger \sigma^{\mu\nu}_{ab} B_{\mu\nu} l_c
+	c_W \tilde{\phi}^\dagger \sigma^{\mu\nu}_{ab} \sigma_i W^i_{\mu\nu} l_c
+	c_\phi \sigma^{\mu\nu}_{ab} (D_{\mu} \tilde{\phi})^\dagger  D_{\nu} l_c \ \Big]  \\
+&	  \text{h.c.}.  
\eea
In this Hamiltonian, the strength of the various interactions is governed by the couplings $c_X$, which are arbitrary and which, taken one-by-one, are only constrained by the fact that they should be small enough for perturbation theory to hold. So, in addition to the scale $\Lambda$, there are six parameters $c_X$ that describe the interactions of a generic singlet field $\psi_{3/2}$ with a mass $m_{3/2}$. 

A few critical comments are in order. Firstly, the higher-spin particles considered in this work are Majorana and thus, for any of their decay modes, also the conjugate modes must be present. Secondly, each interaction term in eq.~\eqref{eq:L-simple} involves the SM leptons and quarks in different ways. Therefore, the interactions can be classified according to the baryon and lepton number created in each interaction, namely by $\Delta B$ and $\Delta L$. In detail, the $c_q$ term creates three quarks and no anti-quarks, implying $\Delta B=1$, while all other terms in \eqref{eq:L-simple} have $\Delta L=1$. However, since $\psi_{3/2}$ is Majorana, no $B$ or $L$ quantum numbers can be assigned to it because the conjugate operators create the configurations with opposite quantum numbers. This is analogous to, but more general than, the case of massive Majorana neutrinos. Thirdly, if only one of the couplings in eq.~\eqref{eq:L-simple} is non-vanishing at a time, there are no severe constraints on their strength. On the other hand, if both the lepton and baryon number violating couplings are present, dangerous processes like proton decay may occur, constraining such combinations. For collider phenomenology purposes, we will keep only one coupling non-vanishing at a time, unless stated otherwise.

We are now in a position to discuss the collider phenomenology of the spin-3/2 particle, its relevant decay modes, present constraints on its mass and couplings and the production cross sections at hadron colliders, as well as the main signatures to which it leads.

%%%%%%%%%%%%%%%%%%%%%%%%%%%%%%%%%%%%%%%%%%%%%%%
\subsection{Decay modes and branching ratios}
%%%%%%%%%%%%%%%%%%%%%%%%%%%%%%%%%%%%%%%%%%%%%%%%

The linear Hamiltonian~\eqref{eq:L-simple} permits the following fermionic decay modes of the spin-3/2 particle $\psi_{3/2}$, adopting, as stated above, the notation of the first family
\bea
& \psi_{3/2} \to udd \ , \ \bar u\bar d\bar d ,\\
& \psi_{3/2} \to e^+e^- \nu_e \ , \  e^+e^- \bar{\nu}_e , d \bar d \nu_e \ , \ d \bar d \bar{\nu}_e \ , \ u \bar d e^- \ , \ \bar u  d e^+ .
%& \psi_{3/2} \to , 
\eea
We have disentangled interactions with $\Delta B=1,\;\Delta L=0$ (first row) and $\Delta B=0,\;\Delta L=1$ (second row) as they should be treated separately. If the masses of the final state fermions can be neglected, which is the case for the first generation, the partial decay widths can be summarized as
\be
    \Gamma(\psi_{3/2}  \to f_1f_2 f_3) =  \frac{\kappa_{f_1 f_2 f_3}}{ 7680 \pi^3}  \, \frac{m_{3/2}^7}{\Lambda^6} \, ,
\ee
where the overall factor $\kappa_{f_1 f_2 f_3}$ depends on the number of quarks in the final state, \eg,
\be\label{eq:kappa-dec}
\kappa_{e^+e^-\nu_e}\! =\! \kappa_{e^+e^-\bar{\nu}_e} \!=\! |c_l|^2,\
\kappa_{udd}\! = \! \kappa_{\bar u \bar d \bar d} \! = \! 3|c_q|^2,
\kappa_{d \bar d \nu_e}\! = \!\kappa_{d \bar d \bar{\nu}_e} \!= \kappa_{u \bar d e^-} \! = \! \kappa_{\bar u  d e^+} \!=\!\frac34 |c_{lq}|^2\,.
\ee

In the $\Delta B=0,\;\Delta L=1$ case, there are also two-body decays into a lepton and a massive gauge or Higgs boson,
\be
\psi_{3/2} \to  W^+ e^- \, ,\  W^- e^+ \, ,  \ Z\nu_e \, , \ Z\bar{\nu}_e \, ,  \  \gamma \nu_e \, , \ \gamma \bar{\nu}_e \, , \  H \nu_e \, , \ H \bar{\nu}_e,. \nonumber
\ee
when $\psi_{3/2}$ is sufficiently heavy. The partial decay widths involving final state gauge bosons are
\bea
\Gamma(\psi_{3/2}  \to  W^{+}e^-) 
    = &\frac{(m_{3/2}^2-M_W^2)^2}{ 768 \pi m_{3/2}^3 \Lambda^6} 
   \Bigg\{c_\phi^2(m_{3/2}^2+6M_W^2)(m_{3/2}^2-M_W^2)^2\\
&   +\frac{16c_\phi c_W}{g_2}M_W^2(m_{3/2}^4+2m_{3/2}^2M_W^2-3M_W^4)\\
&   +\frac{32 c_W^2}{g_2^2}M_W^2(m_{3/2}^4+2m_{3/2}^2 M_W^2+3M_W^4)\Bigg\} \, , \\
\Gamma(\psi_{3/2}   \to  Z\nu_e) 
    = &\frac{(m_{3/2}^2-M_Z^2)^2}{ 768 \pi m_{3/2}^3 \Lambda^6}     
   \Bigg\{\frac{c_\phi^2}{2}(m_{3/2}^2+6M_Z^2)(m_{3/2}^2-M_Z^2)^2 \\ 
&   +\frac{8c_\phi c_Z\cW}{g_2}M_Z^2(m_{3/2}^4+2m_{3/2}^2 M_Z^2-3M_Z^4)\\
&   +\frac{16 c_Z^2 \cos^2\theta_W}{g_2^2}M_Z^2(m_{3/2}^4+2m_{3/2}^2 M_Z^2+3M_Z^4)\Bigg\},
\\
\Gamma(\psi_{3/2}   \to  \gamma \nu_e) 
    = &\frac{c_\gamma^2 v^2}{192\pi \Lambda^6}m_{3/2}^5,
\eea
where we introduced the couplings
\be
    c_\gamma \!\equiv \!-c_B \cW +c_W \sW, \qquad 
    c_Z \! \equiv \!  c_B \sW + c_W \cW , \ 
\ee
and the partial decay width into a Higgs boson and a neutrino is
\be
    \Gamma(\psi_{3/2}   \to  H\nu_e) 
    =   \, \frac{c_\phi^2}{ 1536\pi}  \, \frac{m_{3/2}^7}{\Lambda^6} \, \left(1- \frac{M_H^2}{ m_{3/2}^2} \right)^4 \, . 
\ee
Due to the Majorana nature of $\psi_{3/2}$, decays to final states containing the corresponding antiparticles are also possible and have an equal partial decay width. 

\begin{figure}[t]
\centering
    \includegraphics[width=0.89\linewidth]{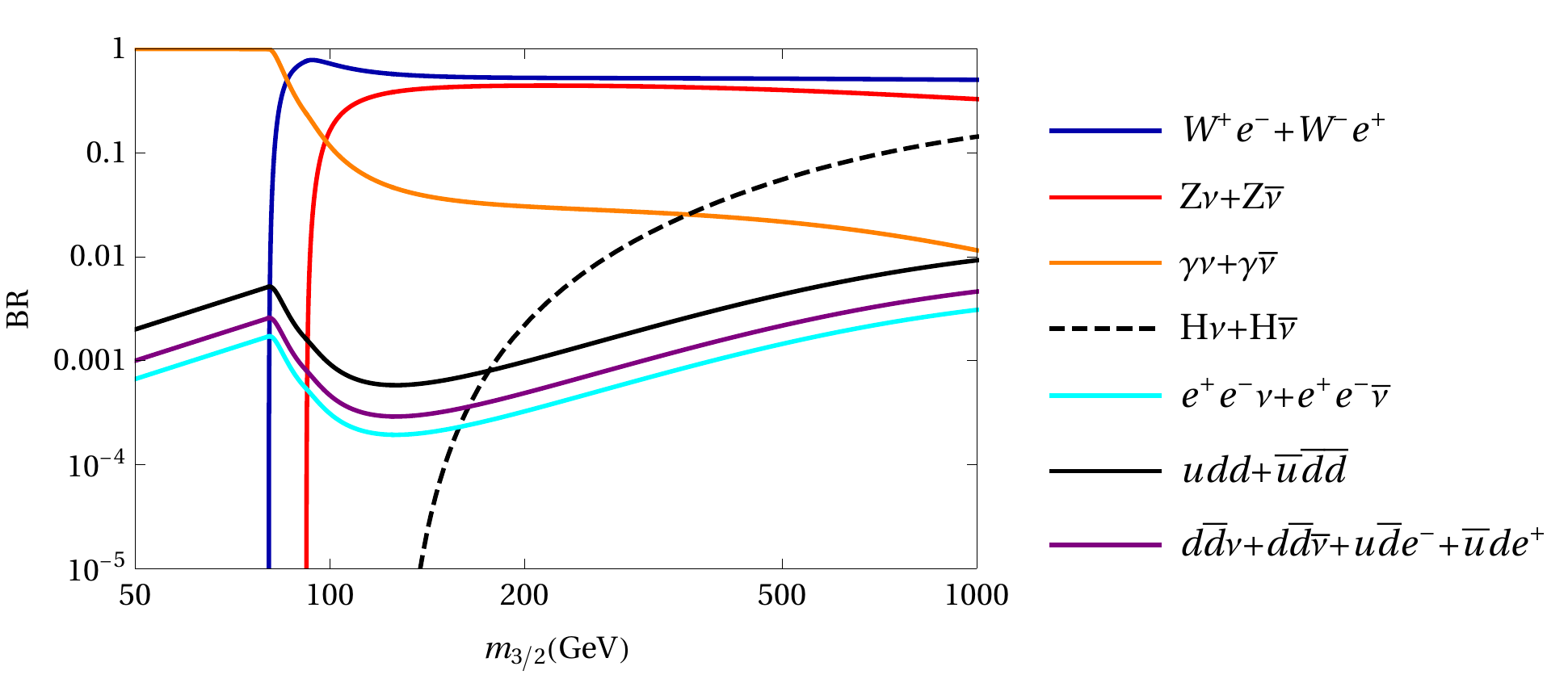}
\caption[]{The decay branching ratios of the $\psi_{3/2}$ states into the various final states as functions of $m_{3/2}$ for $\Lambda=1$~TeV and $c_q=c_l=c_{lq}=c_B=c_W=c_\phi=1$.}
\label{spin-3/2 BR}
\end{figure}

The $\psi_{3/2}$ branching ratios are presented in Fig.~\ref{spin-3/2 BR} as a function of the mass $m_{3/2}$ for the simple case in which all $c_X$ coefficients are equal. As can be seen, the decays into $W^\pm e^\pm$ and $Z\nu+ Z\bar \nu$ final states are by far dominant and have branching ratios that approach the level of 50\% each. The branching ratios for the decays into $H\nu$ and $\gamma\nu$ final states are at the level of a few percent and are comparable for masses around 300 GeV, but they can reach the level of 10\% if the $\psi_{3/2}$ mass is much larger or smaller (a factor $\approx 3$), respectively.  The rates for the decays through the point-like interactions are below the percent level and are larger when the final quark multiplicity is larger according to eq.~(\ref{eq:kappa-dec}).

When the mass of $\psi_{3/2}$ exceeds the electroweak scale, it roughly holds that $\Gamma_{\psi_{3/2}} \propto m_{3/2}^7/\Lambda^6$, thus the total width can grow extremely rapidly with the mass. For instance, taking $\Lambda=1 \, \TeV$ and all coefficients equal to unity, $c_X=1$, we find a total width of about $0.3 \, \MeV$ when $m_{3/2}=200\,\GeV$. However, when increasing the mass to $m_{3/2}=800\,\GeV$ , the total width will grow by several orders of magnitude to about  $0.7 \, \GeV$. On the other hand, due to the strong dependence in $\Lambda$, as long as $m_ {3/2} \lsim \Lambda$, the total width remains extremely small and cannot be resolved experimentally.

%%%%%%%%%%%%%%%%%%%%%%%%%%%%%%%%%%%%%%%%%%%%%%%%%%%%%%%%%%%%%%%%%%%% 
 \subsection{Constraints and production in $e^+e^-$ collisions}
%%%%%%%%%%%%%%%%%%%%%%%%%%%%%%%%%%%%%%%%%%%%%%%%%%%%%%%%%%%%%%%%%%

The most stringent constraints on the new physics scale $\Lambda$ arise from the experimental limits on proton lifetime. Since the spin-3/2 fermion couples to the SM leptons and quarks according to eq.~\eqref{eq:L-simple}, if only one interaction term in this Hamiltonian is present, the constraints on $\psi_{3/2}$ interactions appear only from collider physics and are not stringent, as will be seen below. However, if for instance $c_q$ and $c_l$ are both non-vanishing, $\psi_{3/2}$ mediated processes like $ud \to \bar d e^+e^-\nu$ may give rise to the proton decay to a 4-body final state. Assuming the $\psi_{3/2}$ mass to be close to the cut-off scale, $m_{\psi_{3/2}}\sim \Lambda,$ and taking $c_q=c_l=1,$ our order-of-magnitude estimate for the proton lifetime provides a constraint $\Lambda \gsim {\cal O} (10)$~TeV. This scale is so high that there is little chance that it has been probed so far and, in principle, it should be directly accessible only at future colliders, such as the 100~TeV FCC-hh machine~\cite{Benedikt:2018csr}. Thus, future colliders may be able to probe the parameter space of these higher-spin particles which is currently not constrained by any data.

Nevertheless, the previous strong constraints from proton decay can be simply evaded by requiring that the operators that lead to the $\Delta B=0,\;\Delta L=1$ and $\Delta B=1,\;\Delta L=0$ possibilities do not occur at the same time and, hence, either $c_q$ or $c_l, c_{ql}$ should be zero if one considers the four-fermion operators. In this case, the only constraints on the new states come from collider searches. We will discuss in the following some of these experimental constraints which should have been obtained before the start of the LHC.

The most immediate constraints on the $\Delta B=0,\;\Delta L=1$ spin-3/2 interactions would come from $W,Z$ as well as Higgs boson decays. Indeed, for a mass $m_{3/2} \lesssim M_W, M_Z, M_H$, these particles could decay into a $\psi_{3/2}$ and a lepton,
\be
    W^\pm \to \psi_{3/2} e^\pm \, , \quad 
    Z \to \psi_{3/2}\nu  , \psi_{3/2} \bar \nu \, , \quad 
    H \to \psi_{3/2} \bar \nu , \psi_{3/2} \nu .
    \nonumber
\ee
The partial widths of these decay modes are
\bea
    \Gamma(W^\pm \to \psi_{3/2} e^\pm) 
=&  \frac{(M_W^2-m_{3/2}^2)^2}{ 576 \pi M_W^3 \Lambda^6} \Bigg\{c_\phi^2(m_{3/2}^2+6M_W^2)(M_W^2-m_{3/2}^2)^2
\\
&   +\frac{16c_\phi c_W}{g_2}M_W^2(3M_W^4 -2m_{3/2}^2 M_W^2-m_{3/2}^4)
\\
&   +\frac{32 c_W^2}{g_2^2}M_W^2(3M_W^4 +2m_{3/2}^2 M_W^2+m_{3/2}^4)\Bigg\} \, , 
\\
    \Gamma(Z  \to  \psi_{3/2}\nu_e) = &\frac{(M_Z^2-m_{3/2}^2)^2}{ 576 \pi M_Z^3 \Lambda^6} \Bigg\{\frac{c_\phi^2}{2}(m_{3/2}^2+6M_Z^2)(M_Z^2-m_{3/2}^2)^2\\
&   +\frac{8c_\phi c_Z\cW}{g_2}M_Z^2(3M_Z^4-2m_{3/2}^2M_Z^2-m_{3/2}^4)
\\
&   +\frac{16 c_Z^2 \cos^2\theta_W}{g_2^2}M_Z^2(3M_Z^4+2m_{3/2}^2M_Z^2+m_{3/2}^4)\Bigg\}, 
\\
    \Gamma( H \to \psi_{3/2}  \nu )   
=&  \frac{|c_\phi|^2}{384\pi\Lambda^6}M_H^5m_{3/2}^2 \left(1-\frac{m_{3/2}^2}{M_H^2}\right)^4. 
\eea
Normalized to the total experimentally measured decay widths, $\Gamma_Z^{\rm tot}= 2.4952$ GeV, $\Gamma_W^{\rm tot}= 2.085$ GeV~\cite{Zyla:2020zbs} and $\Gamma_H^{\rm tot}= 4.07$ MeV in the SM~\cite{Djouadi:2018xqq}, and, as before, assuming that $c_\phi=c_W=c_B=1$ and $\Lambda=1$ TeV, the branching ratios for $Z,W$ decays are
\be
    {\rm BR}(Z \to \psi_{3/2}   \nu_e) \simeq 3\times 10^{-6} \, , \qquad 
    {\rm BR}(W^\pm \to \psi_{3/2}  e^\pm) \simeq 2\times 10^{-6} \, ,
\ee
in the favorable case where phase space effects are ignored, \ie, $m_{3/2} \ll M_{W,Z}$. These rates are extremely small and cannot be measured despite of the precise measurement of the massive gauge boson total widths $\Delta \Gamma_Z^{\rm tot}/ \Gamma_Z^{\rm tot} \simeq 0.1\%$ and $\Delta \Gamma_W^{\rm tot}/\Gamma_W^{\rm tot} \simeq 2\%$. In the case of spin-3/2 states from Higgs decays, one has an extremely small partial width $\Gamma(H \to \psi_{3/2} \bar \nu_e) \to 0$ for $m_{3/2} \ll M_{H}$ as it is proportional to $m_{3/2}^2$. 

Nevertheless, for instance, the process $Z \to \psi_{3/2} \nu_e$ with subsequent decays $\psi_{3/2} \to e^+e^- \nu_e,\; d \bar d \nu_e$ or $u \bar d e^-$,\footnote{3-body decays into off-shell intermediate bosons $\psi_{3/2} \to W^* e^\pm \to f\bar f e^\pm$ and $\psi_{3/2} \to Z^* \nu_e \to f\bar f \nu_e$ are also possible. In this case $f$ is an almost massless SM fermion which, for some values of the coefficients $c_X$ could have comparable rates to the direct decays above. The mode $\psi_{3/2} \to H^* \nu_e$, in turn, should be suppressed by the extremely small total decay width of the $H$ boson.} should have been observed in $Z$ decays at LEP1 due to its very high statistics, especially since the signature is rather clean, for $\mathcal{O}(1)$ values of the $c_X$ coefficients and not too large scale $\Lambda$. For the example given above in which ${\rm BR}(Z \to \psi_{3/2} \nu) \approx 3 \times 10^{-6}$, one obtains about 30 clean events for the $10^7$ $Z$ bosons produced at LEP1. Thus, one presumably already has the lower bound $m_{3/2}\lsim M_Z$ for parameter values that allow for the production of the $\psi_{3/2}$ particles at the LHC.

The $\psi_{3/2}$ particle could have been produced at LEP2 for even larger masses than above, as the centre-of-mass energy of the collider slightly exceeded $\sqrt s=200$~GeV. Indeed, taking advantage of the four-point interaction $\psi_{3/2} e^+e^- \bar \nu_e$, the new state can be produced at electron-positron colliders in the process
\be 
    e^+ e^- \to \psi_{3/2} \bar \nu_e \, , \ \psi_{3/2} \nu_e, \nonumber
\ee
with a differential cross section
\bea
    \frac{d\sigma(e^+ e^-\to \psi_{3/2} \bar \nu_e )}{d\cos\theta} 
&   \equiv \frac{d\sigma(e_R^+ e_L^-\to \psi_{3/2} \bar{\nu}_{L}^i)}{d\cos\theta}
+   \frac{d\sigma(e_L^+ e_R^-\to \psi_{3/2} \nu^i)}{d\cos\theta} \\
&=  \frac{ c_l^2 }{48\pi s } \; \frac{ s^3 }{\Lambda^6} \; {\cal F}'(s, m_{3/2} ),
\eea
where $s,t,u$ denote the Mandelstam variables, $\theta$ is the scattering angle and we defined the function
\bea
    {\cal F}'(s, m )  
    \equiv \left(1-\frac{m^2}{s} \right)\, \frac{3stu-m^2(st+su+tu)}{s^3}
\label{eq:Fp-function}
\eea
for future convenience.
%The Mandelstam variables $s,t,u$ which, if $p_1,p_2,p_3$ and $p_4$ are the momenta of the $e^-, e^+, \psi_{3/2}$ and $\bar \nu$ particles, obey the relations $s=(p_1+p_2)^2= 2p_1 \! \cdot \! p_2$, $t= (p_1-p_3)^2 = -\frac{1}{2}(s-m_{3/2}^2)(1-\cos\theta)$ and $u= (p_1-p_4)^2 = -\frac{1}{2}(s-m_{3/2}^2)(1+\cos\theta)$. The previous formula becomes
%\bea
%    {\cal F}'(s, m )  
%   = \frac14 \left(1-\frac{m^2}{s} \right)^2\, \Big[ 3 (1- \cos^2\theta) + 4\frac{m^2}{s}\cos^2\theta (1-\cos^2\theta)\frac{m^4}{s^2} \Big] .
%\label{eq:Fp-function}
%\eea
The corresponding total cross section is
\bea
&   \sigma(e^+ e^-\to \psi_{3/2} \bar \nu_e ) 
%    = \red{\int_{-1}^1 \frac{d\sigma(e^+ e^-\to \psi \bar \nu_e )}{d\cos\theta}}
    =\frac{ c_l^2 }{48\pi s } \; \frac{ s^3 }{\Lambda^6} \; {\cal F}(s, m_{3/2} ),
\eea
where 
\bea\label{eq:F-function}
    {\cal F}(s, m ) \equiv 1-\frac{4m^2}{3s}+\frac{m^8}{3s^4}\, .
\eea

\begin{figure}[t]
\centering
    \includegraphics[width=0.77\linewidth]{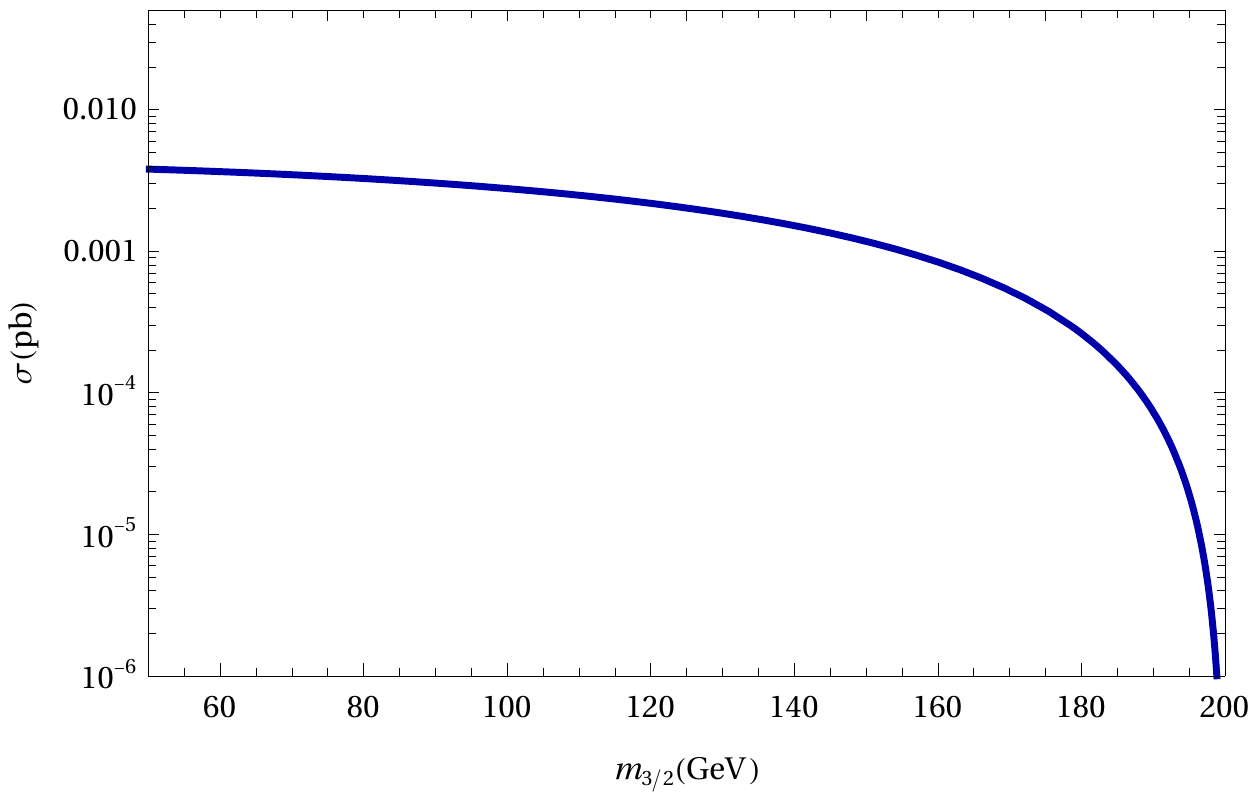}
\caption[]{The production cross section of spin-3/2 particles in the point-like processes $e^+ e^- \to \psi_{3/2} \bar \nu_e + \psi_{3/2} \nu_e$ at LEP2 energies, $\sqrt s= 200$~GeV, as a function of the mass $m_{3/2}$ for the choice of the effective Hamiltonian parameters $\Lambda=1$~TeV and $c_l=1$.}
\label{fig3}
\end{figure} 

The cross section of a single $\psi_{3/2}$ produced in association with a neutrino or an antineutrino through the $\psi_{3/2} \nu e^+ e^-$ contact interaction is depicted Fig.~\ref{fig3} as a function of the mass $m_{3/2}$. We have chosen $\sqrt s= 200$~GeV which typically corresponds to the LEP2 centre-of-mass energy, and $\Lambda=1$~TeV and $c_l=1$. As can be seen, the cross sections are rather small for such parameters, being of the order of a few fb when $m_{3/2}$ is close to 100 GeV. Bearing in mind that the total luminosity at LEP2 was of the order of $\int {\cal L} \approx 100$ pb$^{-1}$, this means that such a parameter set could not be probed. The main reason is that the rate is suppressed by a factor $s^3/\Lambda^6$. in our example we used $\Lambda= 1$ TeV and a c.m. energy $\sqrt s=200$ GeV, corresponding to a suppression factor of $0.2^6 \approx 6 \times 10^{-5}$.

Hence, for masses $m_{3/2} < 200$ GeV, only smaller values of the scale $\Lambda$ and larger coefficients $c_l$, can be excluded at LEP2 via this channel. At future electron-positron colliders, these processes should have more chances to be observed. At the planed FCC-ee option with a c.m. energy of 250 GeV to 350 GeV~\cite{Abada:2019zxq}, the still present large suppression factor will be compensated by the extremely high integrated luminosity, expected to be of the order of a few $ab^{-1}$, \ie, four to five orders of magnitude higher than at LEP2. At planned higher energy electron-positron colliders such as the CLIC machine at CERN with an expected c.m. energy in the TeV range and above, the factor $s^3/\Lambda^6$ ceases to be penalizing if the scale $\Lambda$ is not too high.

The mass of the $\psi_{3/2}$ particle can be fully reconstructed in the considered process by looking at the decays $\psi_{3/2} \to W e^\pm \to q \bar q e^\pm$ and eventually also $\psi_{3/2} \to u \bar d e^-$ as there is no missing energy involved. Moreover, consider the decay $\psi_{3/2} \to Z \nu_e \to e^+e^-\nu_e$ and eventually the direct and more rare decays $\psi_{3/2} \to e^+e^-\nu_e$. They generate the same topology as the pair production of selectrons, the spin--zero superpartners of the electron in the minimal supersymmetric extension of the SM (MSSM), with the selectrons decaying into an electron and the lightest spin-1/2 neutralino $\chi_1^0$, which is supposed to be stable and escapes detection. The process is thus $\epem \to \tilde e^+ \tilde e^- \to \epem \chi_1^0 \chi_1^0$ for which no event has been observed at LEP2 and the limit $m_{\tilde e} > 107$ GeV has been set~\cite{Zyla:2020zbs}. In our case, as we are dealing with single production of the new state in association with a massless neutrino, this should translate into a limit $m_{3/2}\! \gsim \! 200$~GeV for optimistic values of the effective Hamiltonian parameters. 
 
Finally, there is another process for producing the $\psi_{3/2}$ particle at lepton colliders, namely $e^+ e^-\to \gamma^*, Z^* \to \psi_{3/2} \nu_e$ through the operator with coefficient $c_\phi$. Its cross section is expected to be smaller than the one of contact interaction processes and we will discuss this process only in the LHC context to which we turn next.
 
%%%%%%%%%%%%%%%%%%%%%%%%%%%%%%%%%%%%%%%%%%%%%%%%%%%%%%%%%%%%%%%%%%%%%%%%%%
\subsection{Production at hadron colliders and expectations for the LHC }
%%%%%%%%%%%%%%%%%%%%%%%%%%%%%%%%%%%%%%%%%%%%%%%%%%%%%%%%%%%%%%%%%%%%%%%%%%%

The $\psi_{3/2}$ state can be produced at hadron colliders in a leading order process through the $\Delta B=1$, $\Delta L=0$ interaction which couples it to three quarks. In the first generation, several sub-processes involving right-handed up and down type quarks are contributing at the partonic level,
\be
u_R d_R \to \psi_{3/2} \bar d_R   \, , \ \
\bar{u}_R \bar{d}_R \to \psi_{3/2}  d_R \, , \ \
d_R d_R \to \psi_{3/2} \bar{u}_R   \, , \ \
\bar{d}_R \bar{d}_R \to \psi_{3/2} u_R \, . \nonumber
\ee
These subprocesses have equal partonic cross sections which, in term of the scattering angle $\theta$, take the differential form 
\begin{eqnarray} 
    \frac{d \hat \sigma (u_R d_R \to \psi_{3/2} \bar d_R)}{d\cos\theta}  
    &\!=\!& \frac{d \hat \sigma ( \bar{u}_R \bar{d}_R \to \psi_{3/2}  d_R )}{d\cos\theta}  
    \!=\! \frac{d \hat \sigma ( d_R d_R \to \psi_{3/2} \bar{u}_R) }{d\cos\theta}  
    \!=\! \frac{d \hat \sigma (\bar{d}_R \bar{d}_R \to \psi_{3/2} u_R) }{d\cos\theta}  \nonumber 
    \\
    &\!=\!& \frac{c_q^2}{16\pi \hat s }\frac{\hat s ^3 }{\Lambda^6} \, {\cal F}'( \hat s , m_{3/2}) \, ,
\end{eqnarray}
where $\sqrt{\hat s}$ is the partonic centre-of-mass energy and ${\cal F}'$ was given in eq.~\eqref{eq:Fp-function}. Integrating over the scattering angle, the total cross section for each partonic process simply reads
 \be
  \hat \sigma_{i} (q_1 q_2 \to \psi_{3/2} q_3)= \frac{c_q^2}{16\pi \hat s }\frac{\hat s ^3 }{\Lambda^6} \, {\cal F}( \hat s , m_{3/2}) \, ,
\label{eq:sigma-q}
 \ee 
with ${\cal F}$ given in eq.~\eqref{eq:F-function}. To obtain the total hadronic cross section, one should convolute the four partonic cross sections $\hat \sigma_i$, with $i=1\!-\!4,$ over the parton structure functions of the corresponding quarks in the initial state, and sum over the four possibilities of the partonic process. In our numerical analysis, the parton structure functions are chosen to be those of the MSTW2008 fit~\cite{Martin:2009bu}.

In the case of $\Delta B=0,\ \Delta L=1$ interactions, there is a similar process as the one above by which $\psi_{3/2}$ can be produced at hadron colliders at leading order. This is enabled by operator with the $c_{lq}$ coefficient that couples $\psi_{3/2}$ to two quarks and a lepton. Similarly to the previous case, there are four possible partonic processes
\be
d_L \bar{d}_R \to \psi_{3/2} \bar{\nu}_L \, , \ \
\bar{d}_L d_R \to \psi_{3/2} \nu_L \, , \ \
u_L \bar{d}_R \to \psi_{3/2}  e_L^+ \, , \ \ 
\bar{u}_L d_R \to \psi_{3/2}  e_L^-  \, . \nonumber
\ee
with the differential partonic cross sections
\begin{eqnarray}
\frac{d \hat \sigma(d_L \bar{d}_R \to \psi_{3/2} \bar{\nu}_L) }{d\cos\theta}
&\!=\!& \frac{d \hat \sigma(\bar{d}_L d_R\to \psi_{3/2} \nu_L) }{d\cos\theta}
\!=\!\frac{d \hat \sigma(u_L \bar{d}_R\to \psi_{3/2} e_L^+) }{d\cos\theta}
\!+\!\frac{d \hat \sigma(\bar{u}_L d_R\to \psi_{3/2} e_L^-) }{d\cos\theta} \nonumber \\
&\!=\!&\frac{c_{lq}^2}{128\pi \hat s }\frac{\hat s^3}{\Lambda^6} {\cal F}' \, , 
\end{eqnarray}
The total cross sections of the individual partonic processes are
\be
 \hat \sigma_i(q_1 \bar q_2 \to \psi_{3/2} \ell ) = \frac{c_{lq}^2}{128\pi \hat s }\frac{\hat s^3}{\Lambda^6} {\cal F} \, .
\label{eq:sigma-lq}
\ee
One then should convolute these partonic rates over the corresponding parton structure functions and sum over the four possibilities for the individual channels to get the total hadronic cross section of the process. 

\begin{figure}[t]
\centering
    \includegraphics[width=0.85\linewidth]{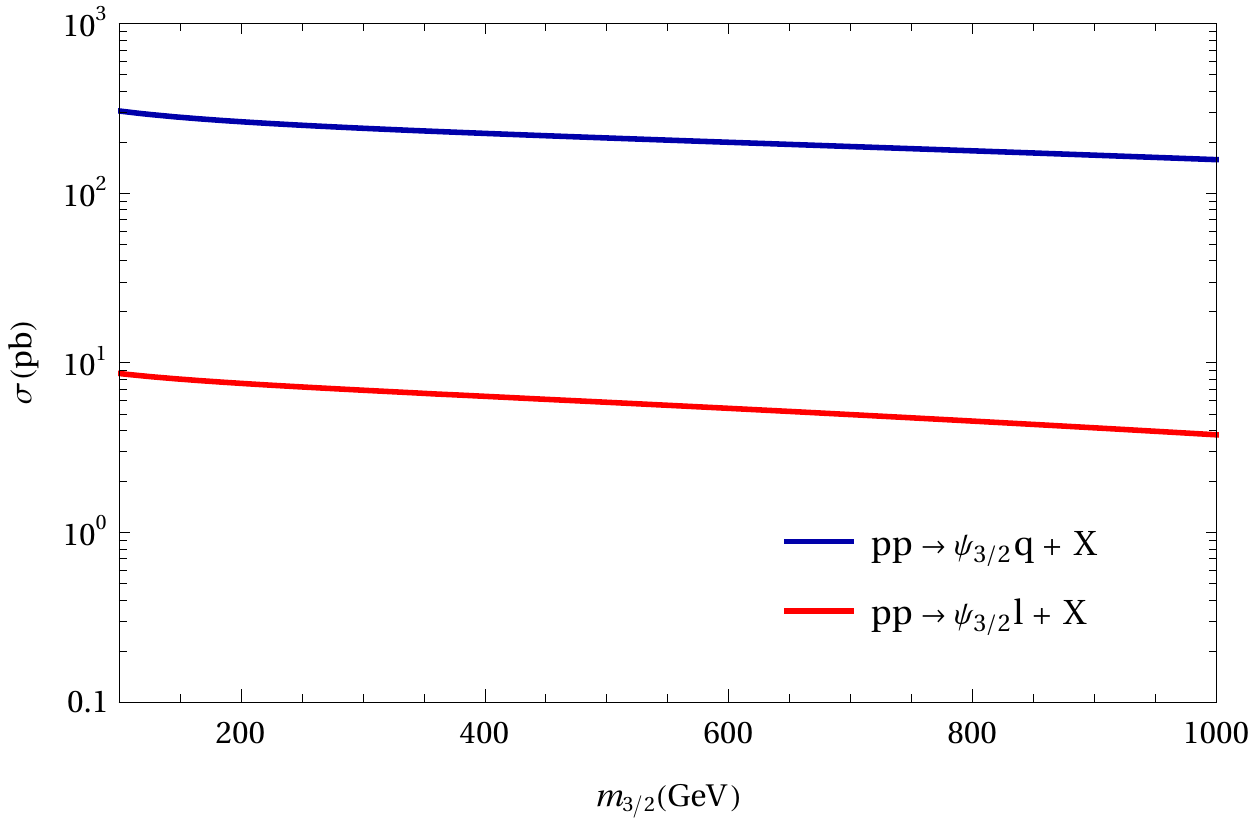} 
\caption[]{The single production cross sections of $\psi_{3/2}$ as a function of $m_{3/2}$ for the two processes $pp\to \psi_{3/2} q + X$ and $pp\to \psi_{3/2} \nu+X$ corresponding to the point-like partonic processes $qq \to \psi_{3/2} q$ and $q\bar q \to \psi_{3/2} \ell$ at the LHC with $\sqrt s= 14$ TeV. The effective Hamiltonian parameters have been set to $\Lambda=1$ TeV and $c_q=c_{lq}=1$.}
\label{fig1}
\end{figure} 

The resulting cross sections for the production of a single $\psi_{3/2}$ at the LHC due to contact interactions are presented in Fig.~\ref{fig1} as a function of the mass $m_{3/2}$ for a collider c.m. energy $\sqrt s= 14$ TeV and effective parameters $\Lambda=1$ TeV and $c_q=c_{lq}=1$. One can easily rescale results for other values of these parameters as the cross sections are proportional to $c^2/\Lambda^{6}$. 

The production cross sections are now fairly large, well above the picobarn level, since the chosen scale of new physics is now of the same order as the partonic c.m. energy such that the suppression $\hat s^3/\Lambda^6$ is not effective anymore. One should also keep in mind that the integrated luminosity which has been collected at the present Run 2 of the LHC is of the order of $\int {\cal L} \approx 140$ fb$^{-1}$ and, hence, is four orders of magnitude higher than the total luminosity obtained at LEP2. Hence, one could collect already a million of $\psi_{3/2}$ events for a production cross section of the order of 10 pb as expected in the process $pp\to \psi_{3/2} l$. The integrated luminosity is expected to significantly increase at the next Run 3 of the collider and even more at the high-luminosity option of the LHC (HL-LHC), where 3 ab$^{-1}$ of data could ultimately be collected. Thus, not only parameters that have never been probed before can be reached at the LHC but also, the sensitivity will benefit from the increase in luminosity and HL-LHC could explore a completely uncovered territory compared to Run~2.

Notice that the two cross sections in Fig.~\ref{fig1} differ by more than an order of magnitude, although $c_q=c_{lq}=1$, that is, they are both induced by contact interactions with equal couplings. The first reason is due to the colour multiplicity producing a factor of 8 in the rate of the process resulting from the coupling $c_q$ compared to the one with the coupling $c_{lq}$, as can bee seen from eqs.~(\ref{eq:sigma-q}--\ref{eq:sigma-lq}). In addition, the cross section in the former case involves only quarks, while the process with the operator of coupling $c_{lq}$ involves a quark and an anti-quark. In proton-proton collisions, the latter cross section is suppressed by the sea-quark parton structure functions. These two features explain the factor of 30 to 40 difference in the two production rates.
 
For the $\psi_{3/2}$ particle that interacts via operators involving gauge and Higgs bosons, there is another process that allows its production at hadron machines: the one occurring through the $s$-channel exchange of a virtual gauge boson in quark-antiquark annihilation. There is also a process with a Higgs boson exchange in the $s$-channel but, because the Higgs boson couples extremely weakly to the first generation quarks, this mode gives negligible cross sections. 

First, there are the neutral current processes involving photon and $Z$--boson exchange 
\be
    q \bar q \to \gamma^*,Z^* \to \psi_{3/2} \nu\, , \psi_{3/2}\bar \nu  \,, \nonumber
\ee
with a neutrino or an antineutrino in the final state. There are also the charged current processes with $W$--exchange, leading to a charged lepton in the final state,
\be
    q \bar q' \to W^{\pm *} \to \psi_{3/2} e^\pm \ . \nonumber
\ee

In the neutral current channel, the differential production cross section of $\psi_{3/2}$, when summing over the $L,R$ helicities  of the initial quarks, reads 
\bea\label{LHC photon Z mediation}
    \frac{d\sigma(q \bar{q} \to \gamma^\ast, Z^\ast\to \psi_{3/2} \nu_e)}{d\cos\theta}
   & =
    \frac{d\sigma(q \bar{q} \to \gamma^\ast, Z^\ast\to \psi_{3/2} \bar{\nu}_{e,L})}{d\cos\theta}
    +
     \frac{d\sigma(q \bar{q} \to \gamma^\ast, Z^\ast\to \psi_{3/2} \nu_{e,L})}{d\cos\theta}\\
&=   \frac{1}{32\pi \hat s} \left(1-\frac{m_{3/2}^2}{\hat s} \right) \sum_{\alpha=L,R}
    |\mathcal{M}_{\alpha \alpha}^q|^2,
\eea
where the amplitude squared of the process is
\be
    |\mathcal{M}_{\alpha \alpha}|^2 
=   \frac{4 \pi v^2}{3 \Lambda^6} \, \frac{e^2}{4 \pi}\,   \Bigg\{
    \left|\frac{c_\gamma e_q}{\hat s}+\frac{c_Z g^Z_{q \alpha} }{\hat s-M_Z^2} \right|^2 {\cal F}_1 
+   \frac{ c_\phi^2 g_2^2 (g^Z_{q \alpha})^2}{\cW^2 (\hat s-M_Z^2)^2} {\cal F}_2  \Bigg\} \, {\cal F}_3 .
\ee
where $e_q$ denote the electric charges of quarks, $e_u=2/3$ and $e_d=-1/3$, the $Z$ couplings to $L/R$ quarks are given by
\be
    g^Z_{q \alpha}= \frac{I_{3q_\alpha} - e_q \sin^2 \theta_W}{ \sW \cW},
\ee
with the isospin $I_{3q} = \pm 1/2$ for the left-handed quarks and $I_{3q} =0$ for the right-handed ones, and we introduced the three functions
\be\label{three fs}
    {\cal F}_1 \equiv \hat s(\hat u-m_{3/2}^2) \, , \ \ 
    {\cal F}_2 \equiv \frac{1}{16}\hat u(\hat s-m_{3/2}^2) \, , \ \ 
    {\cal F}_3 \equiv  m^2_{3/2} \hat t+3\hat s\hat u.
\ee

In the charged current process, the $\psi$ production cross section is driven only by the left-handed quarks,
\begin{eqnarray}
\frac{d\sigma(q \bar{q}\to W^{\ast}\to\psi_{3/2} e)}{d\cos\theta}
    &\!=\!& \frac{d\sigma(u_L \bar{d}_L\to W^{+\ast}\to\psi_{3/2} e_L^{+})}{d\cos\theta}
+\frac{d\sigma(d_L \bar{u}_L\to W^{-\ast}\to\psi_{3/2} e_L^{-})}{d\cos\theta} \nonumber\\
    &\!=\!& \frac{1}{32\pi \hat s}\left(1 - \frac{m^2_{3/2} }{\hat s} \right) 
| \mathcal{M}(q_L \bar{q}_L )  |^2 \, ,\label{LHC W mediation}
\end{eqnarray} 
where the amplitude squared is
\be
    |\mathcal{M}(q_L \bar{q}_L )|^2 
=   \frac{g_2^2 v^2}{3 \Lambda^6}  
    \frac{|V_{\rm CKM}^{ud}|^2}{( \hat s-M_W^2)^2} \Big[ c_W^2 {\cal F}_1  + c_\phi^2 {\cal F}_2   \Big] {\cal F}_3.
\ee

\begin{figure}[t]
\centering
 \includegraphics[width=0.85\linewidth]{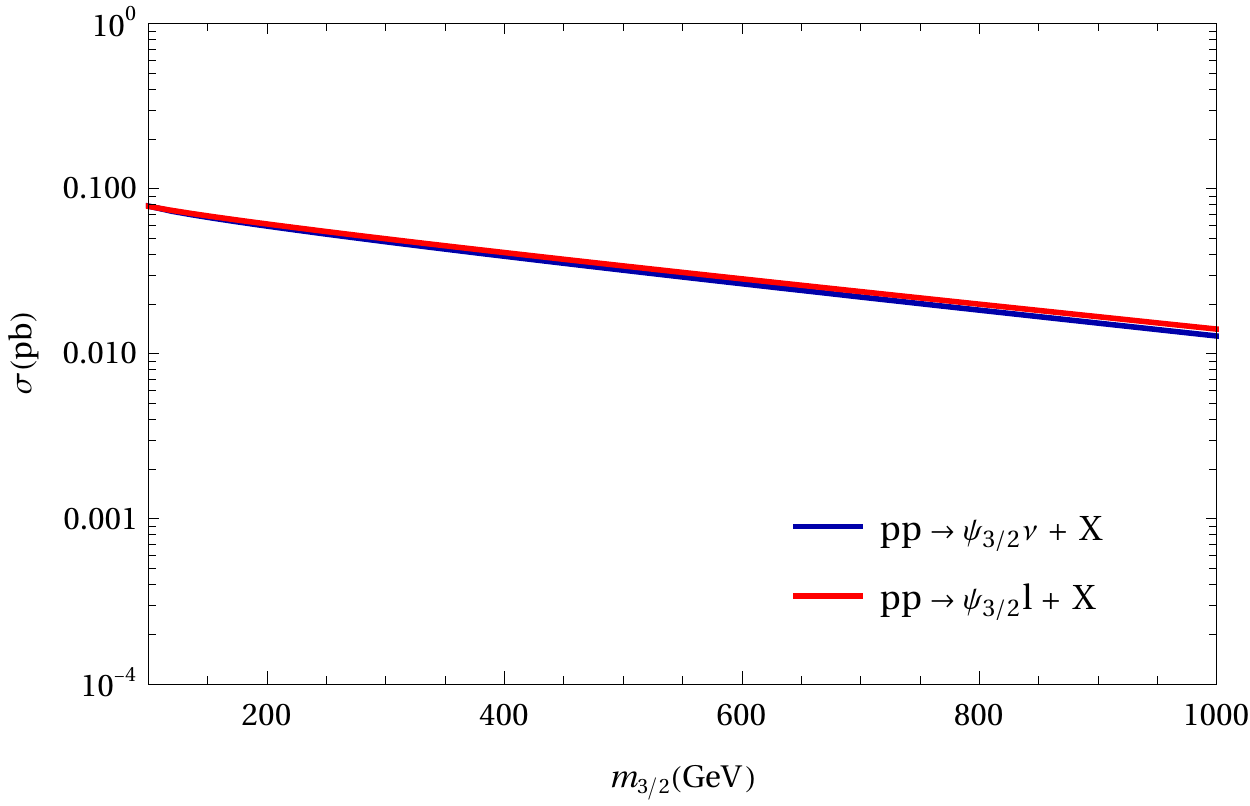} 
\caption[]{The production cross sections of $\psi_{3/2}$ particles at the LHC as a function $m_{3/2}$ for the processes $pp \to \psi_{3/2}\nu +X$ and $pp \to \psi_{3/2} e +X$, corresponding to the parton level processes $q\bar q\to \gamma^\ast, Z^\ast\to\psi_{3/2} \nu$ and $q\bar q\to  W^\ast\to\psi_{3/2} e$ given at eqs. \eqref{LHC photon Z mediation} and \eqref{LHC W mediation} respectively. The choice of parameters is $\sqrt s= 14$ TeV, $\Lambda=1$ TeV and $c_B=c_W=c_\phi=1$.}
\label{fig2}
\end{figure} 

The production cross sections at the LHC for a single $\psi_{3/2}$ state resulting from $s$-channel gauge boson exchange are presented in Fig.~\ref{fig2} for $\sqrt s= 14$ TeV, $\Lambda=1$ TeV and $c_B=c_W=c_\phi=1$. Again, one can easily rescale the results in Fig.~\ref{fig2} for different values of the new physics scale $\Lambda$. The cross sections for producing a neutrino or a charged lepton in the final state are almost identical. A comparison with the cross sections in Fig.~\ref{fig1} shows that the processes mediated by $s$-channel gauge bosons are subdominant compared to the ones induced by the contact interactions, at least two orders of magnitude smaller. This occurs because, in the effective field theory, the two cross sections scale differently with the partonic c.m. energy $\hat s$. Indeed, the former scales as $1/ \hat s$ as is usually the case for $s$-channel gauge boson exchanges, while the latter scales as $\hat s^{2}$ as a result of the contact interaction. This leads to a very striking difference at the high-energy collisions that occur at the LHC. 

%%%%%%%%%%%%%%%%%%%%%%%%%%%
\subsection{Experimental signatures at hadron colliders}
%%%%%%%%%%%%%%%%%%%%%%%%%%

Since our main goal in this paper is to make a general survey of the phenomenology of the new particles with a spin higher than unity, a very detailed account of all the experimental limits that are set by various experiments on their masses and couplings and the expectations in the search for these particles in the future, is clearly beyond our scope. We will nevertheless briefly describe the various signatures of these particles at hadron machines and list the various channels that can be exploited efficiently in their search at the LHC.

Starting with the simpler case of spin-3/2 interactions with $\Delta B\!=\!1$ and $\Delta L\!=\!0$, the unique signature of $\psi_{3/2}$ production at hadron colliders would be the 4-jet final states
\be
   q q \to \psi_{3/2} q \to 4 q \ \ { \Rightarrow }  \ \ pp \to 4j \, . 
\label{topo-quarks}
\ee
In turn, in the case where $\psi_{3/2}$ interacts via the $\Delta B\!=\!0$ and $\Delta L\!=\!1$ operators, the phenomenology is richer and a plethora of final states and, hence, signatures are possible. Focusing first on the operators that involve only point-like fermionic interactions, the various possible final states are 
\be 
\begin{array}{c}
    qq \to \psi_{3/2} \bar \nu \to ee \nu \bar \nu \, , \  qq  \nu\bar \nu \, , \ qq e \nu     \\
    qq \to \psi_{3/2} e         \to eee \nu \, , \ qq e\nu \, ,  \ qq ee   
\end{array}
\ \ { \Rightarrow}  \ \ pp \to ee E_{\rm T}^{\rm mis}, eee E_{\rm T}^{\rm mis},  qq E_{\rm T}^{\rm mis},  qq e E_{\rm T}^{\rm mis}, qq ee \, , 
\label{topo-leptons}
\ee
with $E_{\rm T}^{\rm mis}$ the transverse missing energy carried by the escaping neutrinos. If, instead, one considers the operators in which $\psi_{3/2}$ is coupled to a lepton and a gauge or Higgs boson, the various possible topologies become 
\be
\begin{array}{c}
    qq \to \psi_{3/2} \bar \nu \to We \bar \nu \, , \ Z \nu\bar \nu \, , \ H \nu \bar \nu     \\
    qq \to \psi_{3/2} e        \to Wee \, , \ Z e\nu \, ,  \ H \nu e   
\end{array}
\ \ { \Rightarrow} \ \ pp \to Z E_{\rm T}^{\rm mis}, H E_{\rm T}^{\rm mis}, 
We E_{\rm T}^{\rm mis}, Ze E_{\rm T}^{\rm mis}, He E_{\rm T}^{\rm mis}, Wee \, . 
\label{topo-bosons}
\ee
In the processes above, as we have the subsequent decays $Z \to ll, W \to l\nu$ and $Z,W \to q \bar q$, the initial and final states are just the same as those that occur in the processes of eq.~\eqref{topo-leptons}; if one ignores the difference of magnitude in the branching ratios, only the angular distributions are different, so that both should be combined in principle. However, there are also second and third generation leptonic decays of the massive gauge bosons, which can be used to discriminate between the two possibilities. 
 
As shown above, the vast majority of these processes involve missing energy in the final state. The latter is a typical signature of supersymmetry in models in which the discrete symmetry called $R$-parity is conserved \cite{Farrar:1978xj}. Due to this symmetry, the lightest supersymmetric particle (LSP), generally the lightest neutralino $\chi_1^0$, would be stable and escape detection. Thus, one can use the vast number of supersymmetry searches that have been performed at the LHC and adapt them to our specific case. 

For instance, the signatures $pp \to ee \nu \nu$ and $pp\to qq \nu \nu$ are simply those that appear in the production of right-handed selectrons and first generation quarks that decay directly into the LSP and leptons or light quarks, $pp \to \tilde e_R^+ \tilde e_R^- \to e^+ e^- \chi_1^0 \chi_1^0$ and $pp \to \tilde q \tilde q^* \to q \bar q \chi_1^0 \chi_1^0$.  This is also the case with signatures like $Z/H+ E_{\rm T}^{\rm mis}$ which could be due to the production of the lightest and next-to-lightest neutralino, the latter decaying into the LSP and a gauge or Higgs boson, $q\bar q \to \chi_2^0 \chi_1^0 \to Z/H \chi_1^0 \chi_1^0$. Topologies like $Ze/He/We + E_{\rm T}^{\rm mis}$, as well as those with three and two electron final states produced via four-fermion operators, could be due to slepton pair production and decays through the chain $pp \to \tilde e_R \tilde \nu_L \to e \chi_1^0 \nu_e \chi_2^0 \to e \nu \chi_1^0 \chi_1^0 +Z/H \to Ze+ E_{\rm T}^{\rm mis}$ ($eee+E_{\rm T}^{\rm mis}$ when $Z \to e^+ e^-$) or $pp\to \tilde \nu \tilde \nu \to e^\mp \chi_1^\pm \nu \chi_1^0 \to e^\pm W^\mp + E_{\rm T}^{\rm mis}$ ($ee + E_{\rm T}^{\rm mis}$ when $W \to e\nu)$.

Signatures involving two quarks and charged leptons or neutrinos in the final states are typical of those involving leptoquarks \cite{Djouadi:1989md,Hewett:1987yg} produced in pairs, $pp \to q\bar q, gg \to {\rm LQ} \overline{\rm LQ} \to eq eq, eq q \nu , q\nu q \nu$. The latter signature being also similar to what happens in supersymmetric models for squark pair production, $pp \to \tilde q_R \tilde q_R^* \to q \chi_1^0 \bar q \chi_1^0$ as discussed above. The signature with $Wee$ final states that appear at the end of eq.~\eqref{topo-bosons} would be similar to the one in which a heavy neutrino $N$ is produced in association with an electron through mixing and decays into an electron and a charged $W$ boson, $q\bar q \to W^* \to eN \to eeW$; see for instance Refs.~\cite{delAguila:2007qnc,Djouadi:2016eyy}. Other signatures involving lepton final states can also occur in the production of heavy neutral or charged leptons.

Finally, for the $\Delta B=1$ and $\Delta L=0$ spin-3/2 interactions, only the 4-jet final state topology will be possible, eq.~\eqref{topo-quarks}. This signature is similar to that of squark pair production in $q\bar q$ annihilation or $gg$ fusion, with each squark decaying into two jets in $R$-parity violating supersymmetric processes \cite{Barbier:2004ez}. 

In fact, there is a tight connection between our scenario and the one of $R$-parity violating supersymmetric models. Indeed, the supersymmetric superpotential describing violation of $R= (-1)^{3B+L+2s}$ parity, with $s$ being the spin-number, is~\cite{Dreiner:1997uz} 
\be
    W_{\slash \vspace{-2mm} R} = \lambda_{ijk} L_i L_j \bar E_k+ \lambda'_{ijk} L_iQ_j \bar D_k +\lambda^{"}_{ijk} \bar U_i \bar D_j \bar D_k,
\ee
where $L,E,Q,D,U$ are doublet and singlet superfields involving SM fermions and their spin-0 superpartners. One can see that this superpotential is similar to the one that appears in the first line of the Hamiltonian~\eqref{eq:L-simple} which describes the four--particle interactions of the spin-3/2 particle. Hence, most of the physics of the spin-3/2 particle, as least when the four point-like vertices are concerned, can be described by an $R$-parity violating phenomenon. For instance, the process discussed above, $q\bar q \to \psi_{3/2} q \to qqqq$ is similar to $q\bar q, gg \to \tilde q \tilde q^*$ with the decay $ \tilde q \to q \bar q $ occurring through the $\lambda_{111}^{''} \bar u \bar d \bar d$ operators. 

The four lepton signature resembles the one for slepton pair production with subsequent decays of these into charged leptons or neutrinos via the operator $\lambda_{111} \bar L L \bar e $, leading to $eee\nu$ and $ee\nu \nu$ final states. Also, slepton or squark pair production, in which the pair then decays through the operator $\lambda_{111}' L Q \bar d$ into, respectively, quark pairs and lepton-quark pairs, which finally lead to the $qq ee, qq e \nu$ and $qq \nu \nu$ topologies that also appear in the spin-3/2 case.

Hence, many constraints on squarks and sleptons obtained in $R$-parity violating processes by the ATLAS and CMS collaborations, \eg, Refs.~~\cite{Aad:2019tcc,Sirunyan:2018kzh}, can be used to set constraints on the spin-3/2 particle mass and couplings. As this particle can be produced in association with light SM states, the expectation on the upper limit of the mass $m_{3/2}$ might range from a few TeV, if the couplings to SM particles are order unity, to the level of 100 GeV only, if these couplings are extremely small.

Note, however, that in all the cases discussed above, the same final states have completely different kinematical distributions. For instance, in the decays of $\psi_{3/2}$ into three jets or into three electrons, each of these particles carries a comparable amount of energy which is rather characteristic to a $1 \to 3$ particle decay. For signatures containing neutrino final states, the amount of missing energy should be completely different from the one appearing in supersymmetric scenarios with $R$-parity conservation as, for instance, the supersymmetric particles are produced in pairs leading to the presence of two escaping neutralinos in their decays, while only a single $\psi_{3/2}$ particle is produced in our case, that is, the process would involve only one escaping particle. Hence, care should be taken in adapting the experimental analyses performed in the other scenarios, and it is wiser to conduct some new ones with the specific kinematics of the spin-3/2 particles. 

In addition, in most of the signatures discussed above, the mass of the $\psi_{3/2}$ state cannot be directly reconstructed from the four-momenta of the final particles as they involve missing energy due to the escaping neutrinos. However, there are two important exceptions: {the decays $\psi_{3/2} \to W e^\pm \to q\bar q e^\pm$ as well as $\psi_{3/2} \to ud e^\pm$ do not involve missing energy and the momenta of the two very energetic jets and the electron in the final state combine to form an invariant mass that coincides with the mass of the $\psi_{3/2}$ state.} 

One spectacular signature of the $\psi_{3/2}$ particle would be its production in the $pp\to \psi_{3/2} \nu$ process and its decay through the $\psi_{3/2} \to \gamma \nu $ channel leading to a single and very energetic photon in the final state and a large amount of missing energy. This might be a signature of supersymmetry in some cases, like when the LSP and next-to-LSP neutralino are produced in association, $pp \to \chi_1^0 \chi_2^0$, with a small mass difference that makes the decay $\chi_2^0 \to \chi_1^0 \gamma$ mode rather frequent. This $ \nu \gamma$ spectacular signature has also been discussed long ago in the context of excited neutrinos which can magnetically de-excite into a neutrino and a photon \cite{Boudjema:1989yx}. There is also the mono-Higgs signature, $pp \to \psi_{3/2} \bar \nu \to H\nu \bar \nu$, which could be interesting to exploit if the associated rate is not negligible. 

A more detailed account of all these issues will be postponed to a forthcoming study. 

%%%%%%%%%%%%%%%%%%%%%%%%%%%%%%%%%%%%%%%%%%%%%%%%%%%%%%%%%%%%%%%%%%%%%%%%%%%%%
\section{Spin-2 particles}
\label{sec:spin-2}
%%%%%%%%%%%%%%%%%%%%%%%%%%%%%%%%%%%%%%%%%%%%%%%%%%%%%%%%%%%%%%%%%%%%%%%%%%%%%

%%%%%%%%%%%%%%%%%%%%%%%%%%
\subsection{Interactions} 
%%%%%%%%%%%%%%%%%%%%%%%%%%

The lowest dimension of operators linear in the spin-2 field $\psi_2$ is also 7. For a SM singlet particle, 
$\psi^{abcd}_2$ where $a,b,c,d$ are two-spinor indices, we then have the following effective Hamiltonian 
\bea
	-\mathcal{H}_{\text{linear}}
&	=
	\frac{1}{\Lambda^3} \psi^{abcd}_2 \Big[
	c_B \sigma^{\mu\nu}_{ab} \sigma^{\rho\lambda}_{cd} B_{\mu\nu} B_{\rho\lambda} 
	+ c_W \sigma^{\mu\nu}_{ab} \sigma^{\rho\lambda}_{cd} W_{i\mu\nu} W^i_{\rho\lambda} 
	+ c_G \sigma^{\mu\nu}_{ab} \sigma^{\rho\lambda}_{cd} G_{A\mu\nu} G^A_{\rho\lambda}
	\Big] + \text{h.c.}, 
	\label{eq:psi2}
\eea
where $B_{\mu \nu}, W_{\mu \nu}$ and $G_{\mu \nu}$ are the ${\rm U(1)_Y}$, ${\rm SU(2)_L}$ and ${\rm SU(3)_C}$ field strengths, respectively, and $A$ is a colour-octet index. The coefficients $c_B$, $c_W$ and $c_G$ are in principle arbitrary. We could chose them to be equal as was the case in the spin-3/2 discussion. However, most probably, if they originate from gauge interactions in the ultraviolet regime, they could eventually have the same magnitude as, respectively, the electroweak couplings $g_1$ and $g_2$ and the strong coupling $g_3$ such that $c_G \gg c_B, c_W$. This is the assumption that we will make here, $c_B=g_1(M_Z)$, $c_W=g_2(M_Z)$ and $c_G=g_s(M_Z)$. Note that, unlike for the spin-2 particles in the $(1,1)$ representation (\eg, gravitons), linear interactions with fermions and scalars are absent at the leading order in Eq.~\eqref{eq:psi2} due to Lorentz symmetry.

From this Hamiltonian, one can see that the spin-2 particle $\psi_2$ will couple only to gluons and electroweak gauge bosons and that there are no couplings to fermions nor couplings to the Higgs boson at this order. One could immediately ask whether the terms in eq.~\eqref{eq:psi2} resemble the ones of a massive Kaluza-Klein graviton which was widely discussed in the literature in the context of extra space-time dimensional models, in particular, those with large extra dimensions \cite{Antoniadis:1998ig} and Randall-Sundrum \cite{Randall:1999ee}. There are significant differences between these scenarios and ours. In particular: 

\begin{itemize}
\item[--] $\psi_2$ does not couple to fermions and Higgs bosons, while the massive graviton is more democratic and couples to all particles;\vspace*{-2mm} 
 
\item[--] unlike in extra dimensional models, the $\psi_2$ couplings to gauge bosons are not universal -- the coefficients $c_B, c_W, c_G$ are free parameters and can be different. A practical consequence is that the $\psi_2 Z\gamma$ couplings occur for generic coefficients $c_W$, $c_B$ contrary to the case of extra-dimensional theories;\vspace*{-2mm} 

\item[--] the structure of the interaction operators are different. As a consequence, angular distributions are different from the case of Kaluza-Klein gravitons.\vspace*{-2mm} 

\end{itemize}

These differences come from the fact that our field $\psi_2$ is a generic massive spin-2 field with interactions given by eq.~\eqref{eq:psi2}. Thus, it cannot be identified with the massive graviton. Our interactions also differ from the ones in Ref.~\cite{Fabbrichesi:2020jlb}. Details about the Lorentz group representations of our $\psi_2$ and the ones of massive graviton can be found in Ref.~\cite{Criado:2020jkp}. We now proceed to the phenomenological part and study the properties of the $\psi_2$ state, namely its main decay modes and production channels.

\subsection{Decay modes and branching ratios} 

The spin-2 particle will decay dominantly into the following two-body final states, 
\bea
    \psi_2 \to \gamma\gamma, \ ZZ, \ Z\gamma, \ WW, \ gg.  \nonumber
\eea
Due to the non-Abelian nature of the SM gauge bosons, also three- and four-body final states are possible,
\bea
    &\psi_2  \to \gamma WW, \ ZWW, \ 3g, \\
    &\psi_2  \to 4W, \ \gamma\gamma WW, \ ZZWW, \ \gamma Z WW, \ 4g. \nonumber
\eea 
We will ignore these additional modes as they are of higher order and stick to the two-body final states decays. 

The partial widths of these dominant decays are
\begin{align}
&   \Gamma(\psi\to gg) = \frac{c_G^2}{15\pi }\frac{4m_2^7}{\Lambda^6},\\
&   \Gamma(\psi\to \gamma\gamma) = \frac{(c_B \cos^2\theta_W+c_W \sin^2\theta_W)^2}{30 \pi}\frac{m_2^7}{\Lambda^6},\\
&   \Gamma(\psi\to ZZ) = \frac{(c_B \sin^2\theta_W+c_W \cos^2\theta_W)^2}{30\pi}\frac{(m_2^2-4M_Z^2)^{1/2}(m_2^6+2m_2^4 M_Z^2+36m_2^2 M_Z^4)}{\Lambda^6},\\
&   \Gamma(\psi\to W^+ W^-) = \frac{c_W^2}{15\pi }\frac{(m_2^2-4M_W^2)^{1/2}(m_2^6+2m_2^4 M_W^2+36m_2^2 M_W^4)}{\Lambda^6},\\
&   \Gamma(\psi\to Z\gamma) = \frac{(c_B-c_W)^2 \sin^2\theta_W \cos^2\theta_W}{15\pi }\frac{(m_2^2-m_Z^2)^3}{ \Lambda^6}\left(m_2+\frac{3M_Z^2}{m_2}+\frac{6M_Z^4}{m_2^3}\right).
\end{align}

\begin{figure}[t]
\centering
    \includegraphics[width=0.82\linewidth]{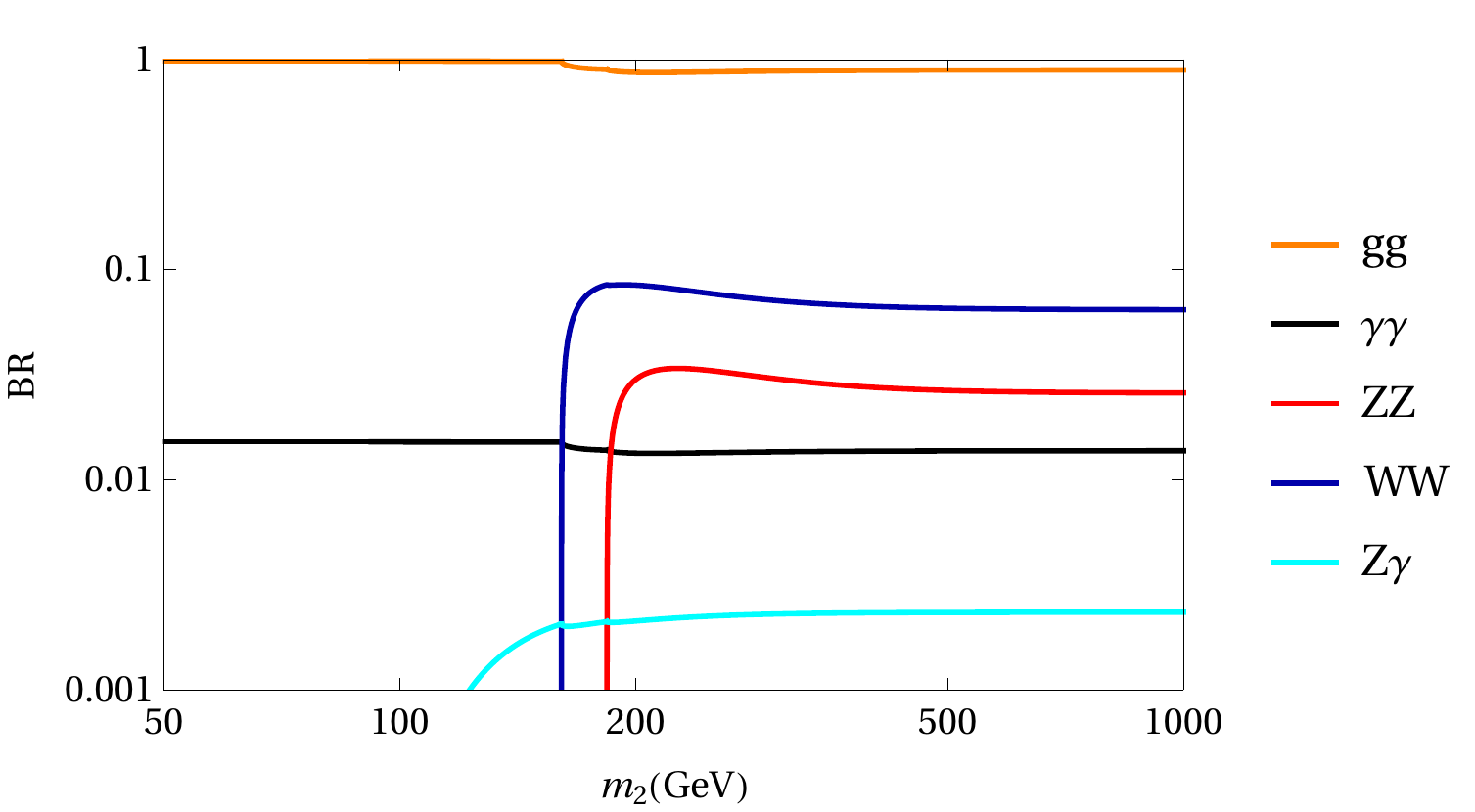}
\caption[]{The branching ratios of $\psi_{2}$ corresponding to different final states as functions of $m_{2}$. The choice of parameters is $\Lambda=1$~TeV and $c_B=g_Y(m_Z), \ c_W=g_2(m_Z), \  c_G=g_s(m_Z)$. }
\label{spin-2 BR}
\end{figure}

The $\psi_2$ branching ratios are presented in Fig.~\ref{spin-2 BR} as a function of the mass $m_2$ for the specific set of coefficients $c_B=g_1(M_Z)=0.36, \ c_W=g_2(M_Z)=0.65$ and $c_G=g_s(M_Z)=1.22$. As expected, the decays into gluons are by far dominant as they involve the strong interaction. The decays $WW, ZZ$ and $\gamma\gamma$ are at a level of 10\% to 2\%, respectively, while the decay into $Z\gamma$ final states stays at the permille level. 

Similarly to the $\psi_{2/3}$ field, the total decay width of a relatively massive $\psi_{2}$ state grows extremely rapidly with the mass since $\Gamma_{\psi_{2}} \propto m_{2}^7/\Lambda^6$. For instance, taking $\Lambda=1 \, \TeV$ and coupling constants $c_B=g_Y(m_Z), \ c_W=g_2(m_Z)$ and $c_G=g_s(m_Z)$, one finds a total width of about  $2 \, \MeV$ when $m_{3/2}=200\,\GeV$. By increasing the mass to $m_{3/2}=800\,\GeV$, the total width will grow by over four orders of magnitude to about  $30 \, \GeV$.

As expected, our results differ from extra-dimensional theories reported in the literature~\cite{Giudice:1998ck,Han:1998sg,Lee:2013bua,Falkowski:2016glr,Dillon:2016fgw,Giddings:2016sfr}, highlighting the point that a generic massive spin-2 field with arbitrary couplings is not necessarily a massive graviton.

\subsection{Main production mechanism and cross sections} 

\begin{figure}[t]
\centering
 \includegraphics[width=0.8\linewidth]{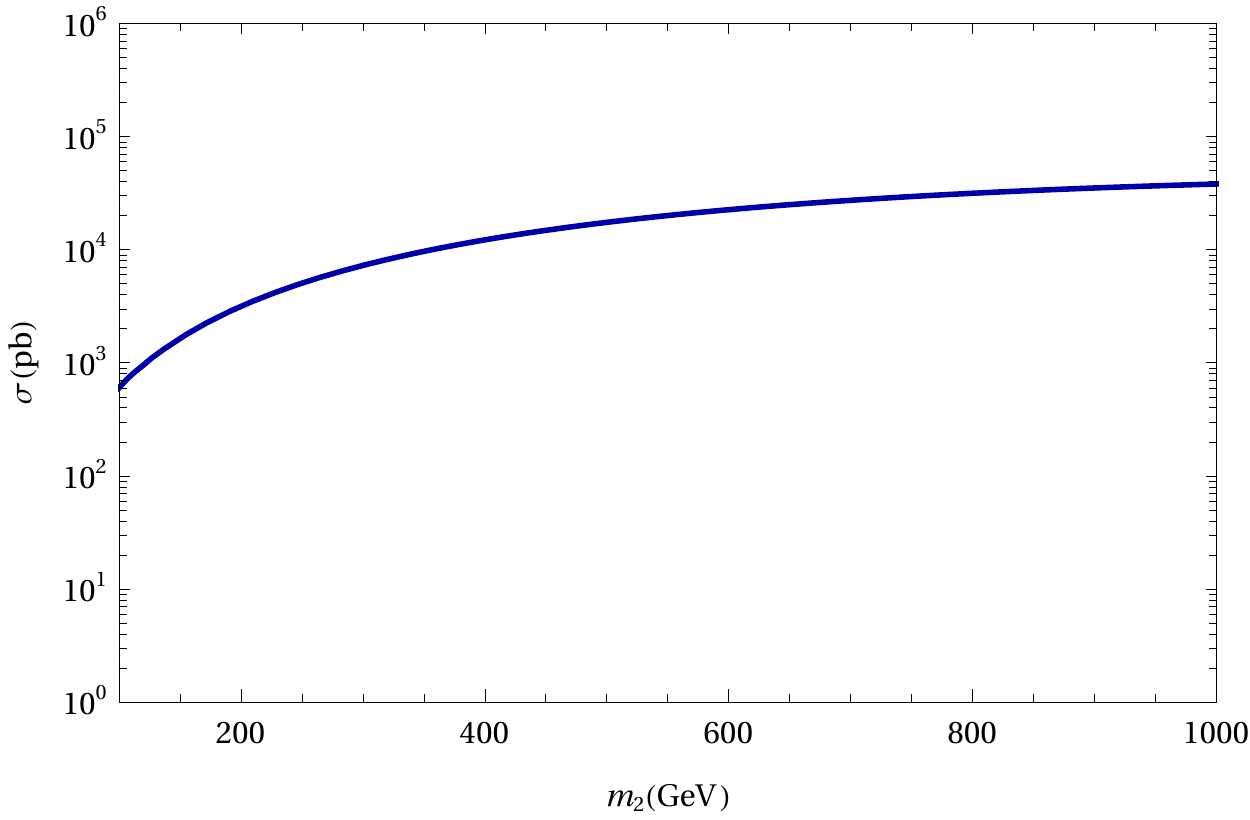} 
\caption[]{The production cross section of $\psi_{2}$ as a function $m_{{2}}$ for the process $pp \to \psi_{2}+X$ corresponding to the partonic process $gg\to \psi_2$ at the LHC with $\sqrt s= 14$ TeV for $\Lambda=1$ TeV and $c_G=1$. }
\label{fig4}
\end{figure}

Because our spin-2 particle couples directly to two gluons but not to quarks, it will mainly be produced in $gg$ fusion, 
\be
     gg\to \psi_2, \nonumber
\ee 
at hadron colliders with the partonic production cross section
\be
    \hat{\sigma}(gg\to \psi_2) = \frac{16\pi c_G^2}{3 }\frac{ m_2^8}{\hat{s}^{3/2}~\Lambda^6}\delta(\sqrt{\hat s}-m_2).
\ee
As usual, to obtain the total cross section, one has to fold this partonic cross section with the gluon luminosities, which are extremely high at high-energies. The latter cross section is shown in Fig.~\ref{fig4} as a function of the spin-2 particle mass $m_2$ for the LHC c.m. energy $\sqrt s= 14$~TeV, and for $\Lambda=1$~TeV and $c_G=1$. Note that according to Refs.~\cite{Kumar:2009nn,Gao:2010bb} in which the higher order corrections to a Kaluza-Klein graviton have been discussed, there might be a $K$-factor of order 2 at the LHC for a TeV scale spin-2 state. Such a $K$-factor has not been included in the plot. As can be seen in Fig.~\ref{fig4}, because we are discussing the single production of a resonance in the $s$-channel, the cross section can be huge. It is at the level of $10^3$ pb at low masses, $m_2 \sim 100$ GeV, but increases to a few times $10^4$ pb when $m_2 \sim 1$ TeV, the steep increase with $m_2^8$ being compensated by the lower probability of finding a gluon in the proton at high resonance masses. With the luminosity of 140 fb$^{-1}$ already collected at the LHC and the additional data to be collected at Run 3, scales of a few TeV could be probed for the chosen $m_2$ range and $c_G$ value. At HL-LHC, scales up to 10 TeV could be probed in this process. 

Note that spin-2 particles can also be produced at photon-photon colliders as $s$-channel resonances in the $\gamma\gamma\to \psi_2$ process. The cross section for this process is similar to that of the two gluon process and is given by
\be
    \sigma(\gamma\gamma\to \psi_2) = \frac{2\pi (c_B\cos^2\theta_W+c_W\sin^2\theta_W)^2}{3 }\frac{ m_2^8}{s^{3/2}~\Lambda^6}\delta(\sqrt{s}-m_2).
\ee
One should then fold this expression with the relevant luminosity of the two photons:  either Weizs\"cker-Williams photons in the usual $e^+e^-$ mode when the photons are simply radiated from the initial beams or Compton backscattered photons from high-power laser beams in the $\gamma\gamma$ option of future linear colliders. 
In addition, one has higher order processes at electron-positron and hadron machines from vector boson and photon fusion, $WW, ZZ, Z\gamma, \gamma\gamma \to \psi_2$ with the former being dominant as usual (because the charged current couplings are larger than the neutral current ones, in general). If only the transverse components of the gauge bosons contribute to the production of the spin-2 object, the rates are expected to be tiny as the luminosity for transverse $W,Z$ bosons is small at high energies, much smaller than the one for the longitudinal component. We shall ignore these higher order processes at the moment.
 
The resonant production will, of course, be followed by the decays of $\psi_2$ into two gauge bosons. Hence, the main topologies to be searched for at proton colliders such as the LHC will be 
\be
    gg \to \psi_2  \to gg, WW, ZZ, Z\gamma, \gamma\gamma. \nonumber
\ee 
%for which, as noted above, the analytic results are pretty simple and, in the case of gravitons, they have been given in papers~\cite{Giudice:1998ck,Bijnens:2001gh}. The rates depend only on $m_{2}$ and $c_G/\Lambda$, plus eventually the parameters $c_B, c_W$ if all final states cannot be reconstructed. $m_{2}$ can be measured with the invariant jet-jet final state masses. Thus if the $c_i$'s are fixed, the only free parameter would be $c_G/\Lambda$, which can be potentially measured at the LHC or at the FCC-hh.
These final states have been searched for at the LHC, in particular in the context of the notorious 750~GeV two-photon resonance that was thought to be observed at the early stage of the Run 2 LHC but which turned out to be a statistical fluctuation. The $gg$ final state is expected to be the dominant one, but one should also focus on the $VV$ ones as they are much cleaner. In particular, the $\gamma\gamma$ signature should be the best as, in the massive bosonic modes, the cleanest final states are those involving charged leptons ($e,\mu$) or neutrinos (missing energy) which are penalized by the small branching ratios as $W,Z$ bosons dominantly decay into $q\bar q$. Hence, the best and most efficient detection signal might be $gg\to \gamma\gamma$. This signature is very clean and has been discussed at length in the literature; see, for instance, Ref.~\cite{Englert:2013opa,Carmona:2016jhr}. 

Finally, we should note that in the experimental signatures, there are also some similarities with the production of spin-1 new neutral bosons $Z'$ and Kaluza-Klein gluons $g_{KK}$. The production modes should be in both cases due to $q\bar q \to Z', g_{KK}$. However, $gg\to g_{KK}$ can also come from the anti-symmetric part of the triple gluon vertex, which invalidates Furry theorem that forbids on-shell 3 vector vertices. So, we have the same initial topology. The $Z'$ will mainly decay into $q\bar q$ (the $WW$ state has a low rate and there are no $ZZ, Z\gamma,\gamma\gamma$ final states). This cannot be discriminated from gluons, except if one looks at angular distributions. But new $Z'$ bosons have direct decays into lepton pairs~\cite{Dittmar:2003ir}, and $g_{KK}$ decays mostly into heavy quark pairs~\cite{Djouadi:2007eg,Allanach:2009vz}. Therefore, one should easily discriminate between these scenarios, even without studying the angular distributions.

%%%%%%%%%%%%%%%%%%%%%%%%%%%%%%%%%%%%%%%%%%%%%%%%%%%%%%%%%%%%%%%%%%%%%%%%%%%%%
\section{Spin-5/2 particles}
\label{sec:spin-5/2}
%%%%%%%%%%%%%%%%%%%%%%%%%%%%%%%%%%%%%%%%%%%%%%%%%%%%%%%%%%%%%%%%%%%%%%%%%%%%%

Turning to spin-5/2 particles, we will simply list here all their possible effective interactions and only briefly describe their collider phenomenology without going into details and performing a numerical analysis. The lowest dimension of the operators linear in a singlet spin-5/2 field $\psi_{5/2}$ is 10. 

The effective Hamiltonian at this order is
\begin{align}
    -\mathcal{H}_{\text{int}}
    &=
    \frac{1}{\Lambda^6} \psi^{abcde}_{5/2} \Big[
    \nonumber \\
    &
    (c^{(1)}_{l})_{ijk} (D_a^{\dot{a}} L^T_{Lbi}) \epsilon (D_{c\dot{a}} L_{Ldj}) e^*_{Rek}
    + (c^{(2)}_{l})_{ijk} (D_a^{\dot{a}} L^T_{Lbi}) \epsilon L_{Ldj} (D_{c\dot{a}} e^*_{Rek})
    \nonumber \\
    &
    + (c^{(1)}_{q})_{ijk} \epsilon_{IJK} (D_a^{\dot{a}} d^*_{RbIi}) (D_{c\dot{a}} d^*_{RdJj}) u^*_{ReKk}
    + (c^{(2)}_{q})_{ijk} \epsilon_{IJK} (D_a^{\dot{a}} d^*_{RbIi}) d^*_{RdJj} (D_{c\dot{a}} u^*_{ReKk})
    \nonumber \\
    &
    + (c^{(1)}_{ql})_{ijk} (D_a^{\dot{a}} Q^T_{LbIi}) \epsilon (D_{c\dot{a}} L_{Ldj}) d^*_{ReIk}
    + (c^{(2)}_{ql})_{ijk} (D_a^{\dot{a}} Q^T_{LbIi}) \epsilon L_{Ldj} (D_{c\dot{a}} d^*_{ReIk})
    \nonumber \\
    &
    + (c^{(3)}_{ql})_{ijk} Q^T_{LbIi} \epsilon (D_a^{\dot{a}} L_{Ldj}) (D_{c\dot{a}} d^*_{ReIk})
    \nonumber \\
    &
    + (c^{(1)}_{qG})_{ijk} \{u^*_{Rai} d^*_{Rbj} d^*_{Rck}\}^A_{\mathbf{S}} \sigma^{\mu\nu}_{de} G^A_{\mu\nu}
    + (c^{(2)}_{qG})_{ijk} \{u^*_{Rai} d^*_{Rbj} d^*_{Rck}\}^A_{\mathbf{A}} \sigma^{\mu\nu}_{de} G^A_{\mu\nu}
    \nonumber \\
    &
    + (c_{qlG})_{ijk} Q^T_{LaIi} \epsilon L_{Lbj} \lambda^A_{IJ} d^*_{RcJk} \sigma^{\mu\nu}_{de} G^A_{\mu\nu}
    + (c_{qlW})_{ijk} Q^T_{LaIi} \epsilon \sigma^A L_{Lbj} d^*_{RcIk} \sigma^{\mu\nu}_{de} W^A_{\mu\nu}
    \nonumber \\
    &  
    + (c_{lW})_{ijk} L^T_{Lai} \epsilon \sigma^A L_{Lbj} e^*_{Rck} 
    \sigma^{\mu\nu}_{de} W^A_{\mu\nu}
    + (c_{lB})_{ijk} L^T_{Lai} \epsilon L_{Lbj} e^*_{Rck} 
    \sigma^{\mu\nu}_{de} B_{\mu\nu}
    \nonumber \\
    &  
    + (c_{qB})_{ijk} \epsilon_{IJK} d^*_{RbIi} d^*_{RdJj} u^*_{ReKk}
    \sigma^{\mu\nu}_{de} B_{\mu\nu}
    + (c_{qlB})_{ijk} Q^T_{LbIi} \epsilon L_{Ldj} d^*_{ReIk}
    \sigma^{\mu\nu}_{de} B_{\mu\nu}
    \nonumber
\end{align}   
\begin{align}
    &  
    + (c_{\phi l D})_{i} \left(D^2_{ab} \widetilde{\phi}\right)^\dagger (D^2_{cd} L_{Lei}) 
    \nonumber \\
    &
    + (c^{(1)}_{\phi l B D})_{i} \left(D^2_{ab} \widetilde{\phi}\right)^\dagger L_{Lci}
    \sigma^{\mu\nu}_{de} B_{\mu\nu}
    + (c^{(2)}_{\phi l B D})_{i}  \widetilde{\phi}^\dagger \left(D^2_{ab} L_{Lci}\right)
    \sigma^{\mu\nu}_{de} B_{\mu\nu}
    \nonumber \\
    &
    + (c^{(1)}_{\phi l W D})_{i} \left(D^2_{ab} \widetilde{\phi}\right)^\dagger \sigma^A L_{Lci}
    \sigma^{\mu\nu}_{de} W^A_{\mu\nu}
    + (c^{(2)}_{\phi l W D})_{i}  \widetilde{\phi}^\dagger \sigma^A \left(D^2_{ab} L_{Lci}\right)
    \sigma^{\mu\nu}_{de} W^A_{\mu\nu}
    \nonumber \\
    &
    + (c_{\phi l B})_{i} \epsilon_{ABC} \widetilde{\phi}^\dagger L_{Lci}
    \sigma^{\mu\nu}_{bc} B_{\mu\nu}
    \sigma^{\rho\sigma}_{de} B_{\rho\sigma}
    + (c_{\phi l BW})_{i} \epsilon_{ABC} \widetilde{\phi}^\dagger \sigma^A L_{Lci}
    \sigma^{\mu\nu}_{bc} B_{\mu\nu}
    \sigma^{\rho\sigma}_{de} W^A_{\rho\sigma}
    \nonumber \\
    &
    + (c_{\phi l W})_{i} \epsilon_{ABC} \widetilde{\phi}^\dagger \sigma^A L_{Lci}
    \sigma^{\mu\nu}_{bc} W^B_{\mu\nu}
    \sigma^{\rho\sigma}_{de} W^C_{\rho\sigma}
    + (c_{\phi l G})_{i} \epsilon_{ABC} \widetilde{\phi}^\dagger L_{Lci}
    \sigma^{\mu\nu}_{bc} G^A_{\mu\nu}
    \sigma^{\rho\sigma}_{de} G^A_{\rho\sigma}
    \Big] + \text{h.c.} \, .
    \label{eq:spin5/2}
\end{align}
The numbers of independent components in flavour space for each of the Wilson coefficient that appear are: 18 for $c^{(1)}_{l}$, $c^{(2)}_{l}$, $c^{(1)}_{q}$, $c^{(2)}_{q}$, $c^{(1)}_{qG}$, $c_{lB}$, $c_{q}$; 27 for $c^{(1)}_{ql}$, $c^{(2)}_{ql}$, $c^{(3)}_{ql}$, $c_{qlG}$, $c_{qlW}$, $c_{qlB}$; 9 for $c^{(2)}_{qG}$, $c_{lW}$; and 3 for $c_{\phi l D}$, $c^{(1)}_{\phi l B D}$, $c^{(2)}_{\phi l B D}$, $c^{(1)}_{\phi l W D}$, $c^{(2)}_{\phi l W D}$, $c_{\phi l B}$, $c_{\phi l WB}$, $c_{\phi l W}$, $c_{\phi l G}$. The product of 3 SU(2) triplets contains 2 octets, which we denote by $\{\}_{\mathbf{S}}$ and $\{\}_{\mathbf{A}}$, with $\{\}_{\mathbf{S}}$ being symmetric in the two last triplets and $\{\}_{\mathbf{A}}$ being anti-symmetric in them. $\lambda^A_{IJ}$ are the Gell-Mann matrices. The rest of the notation is similar to what has been introduced in the spin-3/2 section. 

To deal with the large number of operators and to make the situation more comprehensible, we will simply list the allowed field contents assuming, for simplicity, only the first generation of fermions similarly to the spin-3/2 case,\footnote{Some terms in the list correspond to 3 independent operators.}
\begin{gather*}
  \psi_{5/2} l^2 e^* D^2, \qquad \psi_{5/2} u^* (d^*)^2 D^2, \qquad 3 \times (\psi_{5/2} q d^* l D^2), \\
  \psi_{5/2} u^* (d^*)^2 G, \qquad \psi_{5/2} q d^* l G, \qquad \psi_{5/2} q d^* l W, \\
  \psi_{5/2} l^2 e^* B, \qquad \psi_{5/2} u^* (d^*)^2 B, \qquad \psi_{5/2} q d^* l B, \\
  \psi_{5/2} \phi l D^4, \qquad 3 \times (\psi_{5/2} \phi l W D^2), \qquad 3 \times (\psi_{5/2} \phi l B D^2) \\
  \psi_{5/2} \phi l W^2, \qquad \psi_{5/2} \phi l B W, \qquad \psi_{5/2} \phi l B^2, \qquad \psi_{5/2} \phi l G^2.
\label{eq:spin5/2-simple}
\end{gather*}
There are many possible operators even in the simplified case. As already mentioned, we refrain from a detailed discussion of the phenomenology of the spin-5/2 particles and the computation of the production cross sections and partial decay widths. We consider only the simplest and the most relevant operators and highlight their potential impact on LHC physics. Obviously, the operators involving the strong interactions give the most important contributions in this context: the dominant production processes should involve the gluon-gluon fusion channel and the dominant, and most spectacular decays should involve, respectively, gluons and photons in the final states. From the list of operators displayed in the previous equation, one such example could be
$$ 
    \psi_{5/2} \phi l G^2 \Rightarrow  gg \to \psi_{5/2} \bar \nu \, , \  \psi_{5/2}  \nu ,
$$
for the production mechanism, and also for the most probable final state,
$$ 
    \psi_{5/2} \phi l G^2 \Rightarrow  \psi_{5/2} \to gg \bar \nu , \ gg \nu ,
$$
which means that one gets two jets plus missing energy in the final state. This is the same topology as in the spin-3/2 case with an interaction of
the $ \psi_{3/2}$ field with a quark and lepton where the dominant process was $pp\to d\bar d + u\bar d \to 2j + e^+e^-, 2j+ e\nu, 2j+ \nu \nu$ but where only the last channel might be present. As noted before, the search resembles the one for leptoquarks.

In the case discussed above, one cannot reconstruct the mass of the spin-${5/2}$ particle. To do so, one has to look at other decay modes that do not involve neutrinos and the one that has the highest power of the strong coupling constant and preferentially involves gluons. A possible decay mode could be
$$
    \psi_{5/2}  u (d^*)^2 G \Rightarrow  \psi_{5/2} \to g ud^2 \to 4j,
$$
and the full process would lead to a final state with 4 jets with an invariant mass of $m_2$ plus missing energy. This final state signature has been discussed thoroughly in the context of the experiment as it is a good signature of supersymmetry and it may also test the spin-5/2 particles. Note that as the $ \psi_{5/2}$ interactions are damped by powers of $1/\Lambda^6$ in our case, all decays and productions rates will be proportional to $1/\Lambda^{12}$. This means that for large values of $\Lambda$ the new physics scale they will be extremely small.

%%%%%%%%%%%%%%%%%%%%%%%%%%%%%%%%%%%%%%%%%%%%%%%%%%%%%%%%%%%%%%%%%%%%%%%%%%%%%
\section{Spin-3 particles}
\label{sec:spin-3}
%%%%%%%%%%%%%%%%%%%%%%%%%%%%%%%%%%%%%%%%%%%%%%%%%%%%%%%%%%%%%%%%%%%%%%%%%%%%%

Finally, in the case of the spin-3 field, the lowest dimension of the operators linear in $\psi_3$ is 10 as for $\psi_{5/2}$. However, unlike for $\psi_{5/2}$, the operators linear in $\psi_3$ can have a lower dimension than operators quadratic in $\psi_3$. Spin 3 is the highest spin for which this happens. For a singlet $\psi_3^{abcdef}$ field, the lowest dimensional Hamiltonian is
\bea
	-\mathcal{H}_{\text{linear}}
	=
	\frac{1}{\Lambda^6} \psi^{abcdef}_3
	\sigma^{\mu\nu}_{ab} \sigma^{\rho\sigma}_{cd} \sigma^{\lambda\epsilon}_{ef}
	\Big[ \
&	c_B B_{\mu\nu} B_{\rho\sigma} B_{\lambda\epsilon}   
+	c_{BW} B_{\mu\nu} W_{i\rho\sigma} W^i_{\lambda\epsilon}	\\
+&	c_{BG} B_{\mu\nu} G_{A\rho\sigma} G^A_{\lambda\epsilon}  
+	c_{G} f_{ABC} G^A_{\mu\nu} G^B_{\rho\sigma} G^C_{\lambda\epsilon}
	\ \Big] + \text{h.c.},
	\label{eq:spin3}
\eea
where $f_{ABC}$ are the SU(3) structure constants. Thus, as in the spin-2 case, the spin-3 field does not have a coupling to fermions and couples only to gluons and gauge bosons. As the field strengths involve up to two fields (in the non-Abelian case), there can be interactions of $\psi_3$ with three to six vector fields. However, the more fields one has, the higher is the order in perturbation theory. Thus, the dominant processes will have the minimum number of fields, namely three, and, for simplicity, we will only discuss this option here. 

The most spectacular decay modes of the neutral $\psi_{3}$ state are thus
\be
\psi_3\to  \gamma \gamma \gamma, \gamma \gamma Z, \gamma ZZ, ZZZ, W^+W^-\gamma,W^+W^- 
Z,  
\ee
while the most frequent ones would probably involve gluon jets and could be
\be
    \psi_3\to ggg, gg \gamma, ggZ.
\ee
For instance, the decay rate of $\psi_3$ to three photons is
\be
\Gamma(\psi\to\gamma\gamma\gamma) = \frac{87 c_B^2 \cos^6\theta_W}{11200\pi^3}\frac{m^{13}_3}{\Lambda^{12}}.
\ee
Thus, as in the previous discussion of the spin-2 case, one assumes that the $c_G$ factor for strong interactions is larger than the ones for the weak interactions, $c_B$, $c_{BW},$ and the interferences $c_{BG}$, the main process at hadron colliders will be 
\be 
    gg \to \psi_3 g, \nonumber
\ee
followed~by
\be 
    gg\to  \psi_3 \gamma, \psi_3 Z, \nonumber
\ee
while the main decay modes will be into three gluons, $ \psi_3 \to ggg$ or two gluons and a photon or $Z$ boson, $\psi_3 \to  gg \gamma, ggZ$. 

Keeping only the processes that have more than two (or three) powers of $g_3$ at the amplitude level, one obtains for the various topologies, 
$$ gg \to [4g], [3g\gamma, 3gZ], [2g 2\gamma, 2g2Z, 2g\gamma Z, 2g W^+W^-], \cdots [6\gamma, \cdots ] , $$
with the first bracket being dominant, the second bracket sub-dominant and the third bracket sub-sub-dominant. Of course, the best signal would be $gg\to 4 \gamma$, but it will have a rate that is sub-subleading and very suppressed unless all $c_X$ coefficients are comparable. 

In any case, the dominant process will be either 4 jet events or three jet events with a ``monochromatic" photon. Thus, to the first approximation, one can only consider these two options and ignore all the other possibilities. If a discovery is made, one should try to find the other rare and complicated final states in order to check that it is indeed the signature of a spin-3 particle and derives the various coefficients in the Hamiltonian.

We finally note that at $e^+e^-$ colliders, one could use the process $\gamma \gamma \to \psi_3 +\gamma/Z \to 3g+\gamma/Z$ for production, which might be the dominant one. In the $e^+e^-$ mode, one should rely on vector boson fusion in all channels, $VV \to V'\psi_3$ with $V,V'=\gamma,Z, W$. This also represents the higher order processes that can be probed at hadron colliders too.

%%%%%%%%%%%%%%%%%%%%%%%%%%%%%%%%%%%%%%%%%%%%%
\section{Conclusions}
\label{sec:conc}
%%%%%%%%%%%%%%%%%%%%%%%%%%%%%%%%%%%%%%%%%%%%

We have presented an effective field theory for higher-spin particles that are singlets under the Standard Model gauge group, which involve the lowest order linear operators. The complete set of interactions at that order was explicitly derived for spin-3/2, 2, 5/2 and spin-3 particles and presented in eqs.~\eqref{eq:L-linear}, \eqref{eq:psi2}, \eqref{eq:spin5/2} and \eqref{eq:spin3}, respectively. We have then worked out the most important collider phenomenological features of these particles, mostly at hadron colliders and in particular at the LHC, but also in electron-positron collisions. The partial widths of the principal decay modes and the production cross sections in the main channels have been evaluated for spin-3/2 and spin-2 particles using a formalism introduced earlier, and their implications for constraints on these particles and their search at the LHC have been summarized. In each case, we have discussed the most relevant features and compared the final state experimental signatures with the ones of supersymmetric models or theories of extra dimensions. In the case of spin-5/2 and spin-3 particles, we simply listed the main decay modes and the production mechanisms and highlighted the most striking experimental signatures by which they can be searched for at the LHC and beyond. As the general aim of this work was to pave the way for possible future work on higher-spin particle phenomenology, we have therefore collected the relevant and potentially useful technical material, including Feynman rules and an example of computation, in several appendices. 

The higher-spin particles can have a rich phenomenology and, in addition to the collider aspects discussed in this paper, they can play a potentially important role in other areas of high-energy physics. Higher half--integer spin particles that couple to SM leptons and quarks can, for instance, have some impact on proton decay if some effective operators are simultaneously present. The Majorana nature of the high--spin particles may also have important cosmological consequences, besides those related to dark matter. For instance, if one considers more than one higher-spin particle with different masses and different complex couplings, for example, two different spin-3/2 fermions $\psi_{3/2}^{a},$ $a=1,2,$ interacting according to eq.~\eqref{eq:L-linear}, the possible interference between the complex amplitudes may give rise to direct baryogenesis in the early Universe, very much the same way as in leptogenesis~\cite{Fukugita:1986hr}. In this example, the observed baryon asymmetry of the Universe may not be related to neutrino masses but to the Majorana nature of higher-spin particles. This path is worth exploring.

Finally, while the main focus of the present work has been on collider phenomenology of the hypothetical higher-spin particles, higher-spin resonances do exist in low-energy hadronic physics and in nuclear physics. In the introduction, we reviewed some difficulties related to computing the spin-3/2 resonances described by interacting Rarita-Schwinger fields. Our formulation of the massive interacting spin-3/2 fields is free of those problems and can provide a consistent framework for computing higher-spin nuclear and hadronic physics observables. This aspect is clearly very important and needs a separate and detailed discussion.

\vspace{15mm}
\noindent \textbf{Acknowledgement.}

\noindent This work was supported by the Estonian Research Council grants MOBTTP135, PRG803, MOBTT5, MOBJD323 and MOBTT86, and by the EU through the European Regional Development Fund CoE program TK133 ``The Dark Side of the Universe." J.C.C. is supported by the STFC under grant ST/P001246/1 and A.D. is also supported by the Junta de Andalucia through the Talentia Senior program. 

%%%%%%%%%%%%%%%%%%%%%%%%%%%%%%%%%%%%%%%%  Appendix %%%%%%%%%%%%%%%%%%%%%%%%%%%%%%%%%%%%
\newpage
\appendix
\section{Appendix}
%\setcounter{equation}{0}
%\renewcommand{\theequation}{A.\arabic{equation}}

%%%%%%%%%%%%%%%%%%%%%%%%%%%%%%%%%%%%%%%%%%%%%%%%%%%%%%%%%%%%%%%%%%%%%%%%%%%%%
\subsection*{A1 \; Symmetric multispinor formalism}
\label{app:formalism}
%%%%%%%%%%%%%%%%%%%%%%%%%%%%%%%%%%%%%%%%%%%%%%%%%%%%%%%%%%%%%%%%%%%%%%%%%%%%%

Our notation, first proposed in Ref.~\cite{Criado:2020jkp}, is based on the well-known two-component spinor formalism discussed, \eg, in Refs.~\cite{Dreiner:2008tw, Borodulin:2017pwh}. Undotted indices ($a,b,\ldots$) and dotted indices ($\dot a,\dot b,\ldots$) transform in the $(1/2, 0)$ and $(0,1/2)$ irrep of the Lorentz group, respectively. The indices are raised and lowered with antisymmetric tensors $\epsilon_{ab}$, $\epsilon_{\dot a \dot b}$ with $\epsilon_{12} = -\epsilon^{12} = 1$. When possible, we adapt the convention where undotted (dotted) indices are contracted in descending (ascending) order, \eg~ $t^{a} = \epsilon^{ab}t_{b}$, $t^{\dot a} = t_{\dot a}\epsilon^{\dot a \dot b}$ so that $t^{a} t_{a} = t_{b}\epsilon^{ab}t_{a}$. A pair comprising of a dotted and an undotted spinor index is converted into an vector index $\mu$ as $p_{a\dot a} = p_{\mu}\sigma^\mu_{a\dot{a}}$, $p^{\mu} = \bar{\sigma}^{\mu \dot{a} a}p_{a\dot a}/2$, , where $\sigma^0$ is the identity matrix and $\sigma^i$ with $i=1, 2, 3$ the Pauli matrices, or as $p^{\dot a a} = p_{\mu}\bar\sigma^{\mu\dot{a}a}$, where $\bar{\sigma}^{\mu \dot a b }$ is $\bar\sigma^\mu = (\sigma^{0}, -\sigma^{i})$. It holds that
\be\label{eq:sigma_anticom}
	\sigma^{\mu}_{a\dot a}\bar{\sigma}^{\nu \dot a b } + \sigma^{\nu}_{a\dot a}\bar{\sigma}^{\mu \dot a b } = 2\eta^{\mu\nu} \delta_{a}^{b}.
\ee

Objects in the $(j,0)$ irrep are denoted by $\psi_{(a)} \equiv \psi_{a_1 a_2 \ldots a_{2j}}$, where $(a)$ is a symmetric multispinor index. A multispinor object $t$ is converted into a symmetric multispinor by taking the product of $2j$ copies of $t$ and symmetrizing the indices. For example the momentum $p_{a\dot{a}}$ corresponds to
\be
p_{(a)(\dot{a})} \equiv \frac{1}{(2j)!}\Big[p_{a_1\dot{a}_1}\ldots p_{a_{2j}\dot{a}_{2j}} + \textrm{all permutations of $a_{i}$ and $\dot{a}_j$ }\Big].
\ee
In this way, the $\epsilon_{ab}$ and $\epsilon^{ab}$ symbols are also generalized to the $\epsilon_{(a)(b)}$ and $\epsilon^{(a)(b)}$ symbols that can be used to raise and lower symmetric multispinor indices. The following identities hold
\bea
	\epsilon_{(a)(b)} = (-1)^{2j} \epsilon_{(b)(a)}, \qquad
	\epsilon^{(a)(c)} \epsilon_{(c)(b)} = \delta_{(a)}^{(b)}, \qquad
	\delta_{(a)}^{(a)} = 2j +1.
\eea
The object
\be
    \sigma^{\mu\nu}_{ab} \equiv \frac{i}{4}\left(\sigma^{\mu}_{a\dot a}\bar{\sigma}^{\nu \dot a c} - \sigma^{\nu}_{a\dot a}\bar{\sigma}^{\mu \dot a c}\right)\epsilon_{bc}
\ee
makes a frequent appearance in the Feynman rules. It is symmetric in the two-spinor indices $ab$ and antisymmetric in the Lorentz indices $\mu\nu$ and thus projects rank 2 tensors into their $(1,0)$ subspace.

%%%%%%%%%%%%%%%%%%%%%%%%%%%%%%%%%%%%%%%%%%%%
\subsection*{A2 \; Feynman rules}
\label{app:feyn}
%%%%%%%%%%%%%%%%%%%%%%%%%%%%%%%%%%%%%%%%%%%%

Here we present the Feynman rules for higher spin propagators and vertices with up to 3 legs that arise from a given effective operator. Below $\theta_W$ is the Weinberg angle, $g_Y$, $g_2$, $e = g_2 \sW = g_Y \cW$ are the U$(1)_Y$, SU(2) gauge couplings and the electric charge respectively. All vertices are completely symmetric in the spinor indices.
%m_W = g_2 v/2 = c_W m_Z

%%%%%%%%%%%%%%%%%%%%%%%%%%%%%%%%%%%%%%%%%%%%%%%%%%%%%%%%%%%%%%%%%%%%%%%%%%%%%%%%%%%%%%%%%%%%%
\subsubsection*{Propagators}

\begin{eqnarray}
    \propagatorOne &= i\frac{p_{(a)(\dot{a})}}{p^2-m^2} ,
    \qquad \qquad
    \propagatorTwo = i\frac{p^{(\dot{a})(a)}}{p^2-m^2} ,\\
    \propagatorThree &= i\frac{m^{2j}\delta^{(b)}_{(a)}}{p^2-m^2} ,
    \qquad \qquad
    \propagatorFour = i\frac{m^{2j}\delta^{(\dot{a})}_{(\dot{b})}}{p^2-m^2}.
\end{eqnarray}
\vspace*{6mm}

\subsubsection*{External lines}

\bea\label{external lines}
    \incomingLeft &= u_{(a)}(p,\sigma) ,
    \qquad \qquad
    \incomingRight = v^{\ast}_{(\dot{a}
    )}(p,\sigma) ,\\
    \outgoingLeft &= u^\ast_{(\dot{a})}(p,\sigma) ,
    \qquad \qquad
    \outgoingRight = v_{(a)}(p,\sigma).
\eea
\vspace*{6mm}

\subsubsection*{Completeness relations}

\bea\label{completeness relations}
    \sum_\sigma u_{(a)}(p,\sigma)  u^\ast_{(\dot{a})}(p,\sigma) 
&=   \sum_\sigma v_{(a)}(p,\sigma) v^\ast_{(\dot{a})}(p,\sigma)
=   p_{(a)(\dot{a})},\\
    \sum_\sigma u_{(a)}(p,\sigma) v^{(b)}(p,\sigma)
=  m^{2j}\delta^{(b)}_{(a)}&, \qquad
    \sum_\sigma {u^\ast}^{(\dot a)}(p,\sigma) {v^\ast}_{(\dot b)}(p,\sigma)
=  m^{2j}\delta^{(\dot a)}_{(\dot b)} .
\eea
\vspace*{6mm}

%%%%%%%%%%%%%%%%%%%%%%%%%%%%%%%%%%%%%%%%%%%%%%%%%%%%%%%%%%%%%%%%%%%%%%%%%%%%%%%%%%%%%%%%%%%%%
\subsubsection*{Vertices for spin-3/2 particles}

Here, we present only the vertices with 1) the quartic point-like interaction with three fermions and 2) with up to 3 particles for the interactions with Higgs and gauge bosons which give the dominant interactions; there are also 4 particle vertices of two types compared to the ones below: either one can add a Higgs line and the interaction is suppressed by a power of $v$ or add a gauge boson line and the interaction is suppressed by an additional gauge coupling.

\bea
    \psieLeRnuL &= -i\frac{c^{ee\nu}_{ijk}+c^{ee\nu}_{jik}}{\Lambda^3} \delta^d_a\delta^e_b\delta^f_c ,
\\
    \psiuRdRdR &= i\frac{c_{ijk}^{udd}\epsilon^{IJK}}{\Lambda^3} \delta^d_a\delta^e_b\delta^f_c ,
\\
    \psidLdRnuL &= -i\frac{c_{ijk}^{dd\nu}\delta^{IJ}}{\Lambda^3} \delta^d_a\delta^e_b\delta^f_c ,
\eea
\bea
    \psiuLdReL &= i\frac{c_{ijk}^{ude}\delta^{IJ}}{\Lambda^3} \delta^d_a\delta^e_b\delta^f_c ,
\eea
\bea
    \psihnuL = -i\frac{c_i^\phi}{\Lambda^3}\frac{1}{\sqrt{2}}\sigma^{\mu\nu}_{ab}q_{2\mu} q_{1\nu}\delta^d_c ,
\label{psi to h nu feynman rule}
\eea
\bea
    \psiAnuL &=  \frac{v}{\Lambda^3}\sqrt{2}\left( -c_i^B\cW  +c_i^W \sW  \right)\sigma^{\mu\nu}_{ab} q_{2\mu} \delta^d_c ,
\\
    \psiZnuL &= \frac{v}{\Lambda^3}\Big(-i \frac{e c_i^\phi}{2\sqrt{2}\sW \cW} q_{1\mu}\\
         &\qquad  +\sqrt{2}(c_i^B\sW+c_i^W \cW  )q_{2\mu} \Big)\sigma^{\mu\nu}_{ab} \delta^d_c ,
\\
    \psiWpluseL &=  \frac{v}{\Lambda^3}\left(-i c_i^\phi \frac{g_2}{2}  q_{1\mu} 
+ c_i^W 2   q_{2\mu}\right)\sigma^{\mu\nu}_{ab}\delta^d_c .
\eea

%%%%%%%%%%%%%%%%%%%%%%%%%%%%%%%%%%%%%%%%%%%%%%%%%%%%%%%%%%%%%%%%%%%%%%%%%%%%%%%%%%%%%%%%%%%%%
\subsubsection*{Vertices for spin-2 particles}

\bea
    \SpinTwopsiZZ &= -i\frac{8(c_B \sW^2+c_W \cW^2)}{\Lambda^3} \sigma^{\mu\nu}_{ab}\sigma^{\rho\lambda}_{cd}q_{1\mu} q_{2\rho} ,
\\
    \SpinTwopsiAZ &= i\frac{8(c_B-c_W) \sW \cW}{\Lambda^3}\sigma^{\mu\nu}_{ab}\sigma^{\rho\lambda}_{cd} q_{1\mu} q_{2\rho} ,
\\
    \SpinTwopsiAA &= -i\frac{8(c_B \cW^2+c_W \sW^2)}{\Lambda^3}\sigma^{\mu\nu}_{ab}\sigma^{\rho\lambda}_{cd}q_{1\mu} q_{2\rho} ,
\eea
\bea
    \SpinTwopsiWplusWminus &= -i\frac{8c_W }{\Lambda^3}\sigma^{\mu\nu}_{ab}\sigma^{\rho\lambda}_{cd}q_{1\mu} q_{2\rho} ,
\\
    \SpinTwopsigg &= -i\frac{8c_G}{\Lambda^3}\sigma^{\mu\nu}_{ab}\sigma^{\rho\lambda}_{cd}q_{1\mu} q_{2\rho}\delta_B^A .
\eea

\subsubsection*{Vertices for spin-3 particles}

\bea
    \SpinThreepsiAAA & = -\frac{12\cW(4 c_B \cos^2\theta_W+c_{BW}\sin^2\theta_W)}{\Lambda^6} \sigma^{\mu\nu}_{ab} \sigma^{\rho\sigma}_{cd}\sigma^{\lambda\epsilon}_{ef}q_{1\mu}q_{2\rho}q_{3\lambda},\nonumber
\\
    \SpinThreepsiZZZ & = \frac{12\sW(4 c_B \sin^2\theta_W+c_{BW}\cos^2\theta_W)}{\Lambda^6} \sigma^{\mu\nu}_{ab} \sigma^{\rho\sigma}_{cd}\sigma^{\lambda\epsilon}_{ef}q_{1\mu}q_{2\rho}q_{3\lambda},\nonumber 
\\
    \SpinThreepsiAAZ & = -\frac{8 \sW\cos^2\theta_W(6c_B+c_{BW})}{\Lambda^6} \sigma^{\mu\nu}_{ab} \sigma^{\rho\sigma}_{cd}\sigma^{\lambda\epsilon}_{ef}q_{1\mu}q_{2\rho}q_{3\lambda},
\eea
\bea
    \SpinThreepsiAZZ & = \frac{8 \sin^2\theta_W\cW(6c_B+c_{BW})}{\Lambda^6} \sigma^{\mu\nu}_{ab} \sigma^{\rho\sigma}_{cd}\sigma^{\lambda\epsilon}_{ef}q_{1\mu}q_{2\rho}q_{3\lambda},
\\
    \SpinThreepsiAWW & = -\frac{2\cW c_{BW}}{\Lambda^6} \sigma^{\mu\nu}_{ab} \sigma^{\rho\sigma}_{cd}\sigma^{\lambda\epsilon}_{ef}q_{1\mu}q_{2\rho}q_{3\lambda},
\\
    \SpinThreepsiZWW & = \frac{2\sW c_{BW}}{\Lambda^6} \sigma^{\mu\nu}_{ab} \sigma^{\rho\sigma}_{cd}\sigma^{\lambda\epsilon}_{ef}q_{1\mu}q_{2\rho}q_{3\lambda},
\eea
\bea
    \SpinThreepsiAGG & = -\frac{16\cW c_{BG}}{\Lambda^6} \sigma^{\mu\nu}_{ab} \sigma^{\rho\sigma}_{cd}\sigma^{\lambda\epsilon}_{ef}q_{1\mu}q_{2\rho}q_{3\lambda},
\\
    \SpinThreepsiAGG & = \frac{16\sW c_{BG}}{\Lambda^6} \sigma^{\mu\nu}_{ab} \sigma^{\rho\sigma}_{cd}\sigma^{\lambda\epsilon}_{ef}q_{1\mu}q_{2\rho}q_{3\lambda},
\\
    \SpinThreepsiZGG & = -\frac{16\cW c_{BG}}{\Lambda^6} \sigma^{\mu\nu}_{ab} \sigma^{\rho\sigma}_{cd}\sigma^{\lambda\epsilon}_{ef}q_{1\mu}q_{2\rho}q_{3\lambda},
\\
    \SpinThreepsiGGG & = -\frac{48 c_{G}f_{ABC}}{\Lambda^6} \sigma^{\mu\nu}_{ab} \sigma^{\rho\sigma}_{cd}\sigma^{\lambda\epsilon}_{ef}q_{1\mu}q_{2\rho}q_{3\lambda}.
\eea

%%%%%%%%%%%%%%%%%%%%%%%%%%%%%%%%%%%%%%%%%%%%
\subsection*{A3 \; The narrow width approximation}
\label{app:NWA}
%%%%%%%%%%%%%%%%%%%%%%%%%%%%%%%%%%%%%%%%%%%%

Consider a process $A \to B (\psi \to C)$ that can be split it into the sub-processes $A \to B \psi$ and $\psi \to C$ with amplitudes $\Mcal_{1}(s) \equiv \Mcal_{1}^{(a)}v_{(a)}(s) + \Mcal_{1}{}_{(\dot a)} u^{*}{}^{(\dot a)}(s)$, $\Mcal_{2}(s) \equiv u^{(b)}(s) \Mcal_{2}{}_{(b)} + v^{*}_{(\dot b)}(s) \Mcal_{2}^{(\dot b)} $. The symbols $A,B,C$ denote external states that may contain multiple particles and $s$ is the spin of $\psi$. 

The total amplitude is given by
\bea
	\Mcal 
&	= \frac{\Mcal_{1}^{(a)} i p_{(a)(\dot b)} \Mcal_{2}^{(\dot b)} }{p^2 - m^2 + i m \Gamma} 
	+ \frac{\Mcal_{1}{}_{(\dot a)} i p^{(\dot a)(b)} \Mcal_{2}{}_{(b)} }{p^2 - m^2 + i m \Gamma} \\
&	+ \frac{ \Mcal_{1}{}_{(\dot a)} i m^{2j} \delta^{(\dot a)}{}_{(\dot b)} \Mcal_{2}^{(\dot b)} }{p^2 - m^2 + i m \Gamma} 
	+ \frac{\Mcal_{1}^{(a)} i m^{2j} \delta_{(a)}{}^{(b)} \Mcal_{2}{}_{(b)} }{p^2 - m^2 + i m \Gamma} ,
\eea
where $\Gamma$ and $m$ are the width and mass of $\psi$, respectively. The narrow width approximation works in the limit $\Gamma/M \to 0$ in which case $| p^2 - m^2 - i m \Gamma |^{-2} = \pi/(m \Gamma) \delta(p^2 - m^2)$ effectively putting the intermediate $\psi$ on-shell. The spin sums then imply that $|\Mcal |^2 = \pi/(m \Gamma) \delta(p^2 - m^2) |\sum_{s}\Mcal_{1}(s)\Mcal_{2}(s)|^2$
%\bea
%	\Mcal 
%&	\sim \frac{\Mcal_{1}\Mcal_{2} }{p^2 - m^2 + i m \Gamma}
%	|\Mcal |^2
%&	= \frac{\pi }{m \Gamma} \delta(p^2 - m^2) |\Mcal_{1}|^2 |\Mcal_{2}|^2
%\eea
and thus the cross section decomposes as expected
\bea
	\td \sigma_{A \to B (\psi \to C)}
%	=  \frac{1}{2s}|\Mcal |^2 \td \Pi_{BC}
%&	=  \frac{1}{2s} \delta(p^2 - m^2) |\Mcal_{1}|^2  \frac{}{2m \Gamma}  |\Mcal_{2}|^2 (2\pi) \delta(p^2 - m^2) \td \Pi_{BC} \\
%&	=  \frac{1}{2s} |\Mcal_{1}|^2  \td \Pi_{B\psi} \frac{1}{2m \Gamma}  |\Mcal_{2}|^2 \td \Pi_{C} \\
&	=  \sum_{s_1,s_2} \td \sigma^{s_1s_2}_{A \to  B \psi_{s} } {\td \Gamma^{s_1s_2}_{\psi_{s} \to C}}/{\Gamma}.
\eea
When summing over all final state spins in $C$ and integrating over the $C$ phase space, then $\Gamma^{s_1s_2}_{\psi \to C} = \delta_{s_1s_2} \Gamma_{\psi \to C}$. The cross section for the complete process is thus obtained by performing the spin sum in $\td s_{A \to  B \psi}$ and multiplying the latter by the branching ratio of $\psi \to C$,
\bea
	\td \sigma_{A \to  B (\psi \to C)}
&	=  \td \sigma_{A \to  B \psi } \, {\rm BR}_{\psi \to C}.
\eea

%%%%%%%%%%%%%%%%%%%%%%%%%%%%%%%%%%%%%%%%%%%%
\subsection*{A4 \; Calculation example: spin 3/2 decay}
\label{app:example}
%%%%%%%%%%%%%%%%%%%%%%%%%%%%%%%%%%%%%%%%%%%%

To clarify our formalism, we present here explicit calculation of spin-3/2 particle decay to Higgs and neutrino or anti-neutrino: $\psi_{3/2}(p)\to H(q_2)\nu(q_1)$ and $\psi_{3/2}(p)\to H(q_2) \bar \nu(q_1)$. The contributing diagrams are 
\be
\psiTohnuExample , \quad\quad\quad \psiTohbarnuExample .
\ee
The final states are different and therefore there is no interference between these diagrams. We use two-spinor Feynman rules for the standard model fermions presented in \cite{Dreiner:2008tw}. By using the Feynman rules in \cite{Dreiner:2008tw} and in eqs.~\eqref{external lines} and \eqref{psi to h nu feynman rule} one can write the Feynman amplitude corresponding to the left diagram:
\be
\mathcal{M}_{\psi_{3/2}\to H\bar\nu}
=
-i\frac{c_\phi}{\sqrt{2}\Lambda^3}u^{abc}(p,\sigma)~\sigma^{\mu\nu}_{ab} ~ q_{2\mu}q_{1\nu}~\delta^d_c ~ y_d(q_1,s),
\ee
where $y_d(q_1,s)$ is the wave function corresponding to neutrino. 
The spin-averaged amplitude squared is:
\bea
|\mathcal{M}_{\psi_{3/2}\to H\bar\nu}|^2 
= & \frac{1}{2j+1}\frac{|c_\phi|^2}{2\Lambda^6}
    \left(\sum_{\sigma}u^{\ast \dot a \dot b \dot c}(p,\sigma)u^{def}(p,\sigma)\right)
    \left(\sum_s y^\ast_{\dot c}(q_1,s)y_f(q_1,s)\right)\times\\
&   \times\bar{\sigma}^{\mu\nu}_{\dot b\dot a} \sigma^{\alpha\beta}_{de} q_{2\mu}q_{2\alpha}q_{1\nu}q_{1\beta}.
\eea
This can be simplified by using Pauli-matrix identities:
\be
\sigma^{\mu\nu}_{ab}=\frac{i}{2}\eta^{\mu\nu}\epsilon_{ab}+\frac{i}{2}(\sigma^\mu\bar \sigma^\nu)_{ab}\quad \textrm{and} \quad 
\bar\sigma^{\mu\nu}_{\dot a \dot b}=\frac{i}{2}\eta^{\mu\nu}\epsilon_{\dot a\dot b}+\frac{i}{2}(\bar\sigma^\mu\sigma^\nu)_{\dot a \dot b},
\ee
where the terms proportional to antisymmetric $\epsilon$ vanish when its indices are contracted with the symmetric  $u^{\ast \dot a \dot b \dot c}(p,\sigma)$ and $u^{def}(p,\sigma)$. By using the completeness relation in eq.~\eqref{completeness relations} and in \cite{Dreiner:2008tw} the amplitude squared becomes:
\bea
|\mathcal{M}_{\psi_{3/2}\to H\bar\nu}|^2 
=  -\frac{1}{2j+1}\frac{|c_\phi|^2}{48\Lambda^6}  \Big[
\bar \sigma^{\dot a d}_\mu \bar \sigma^{\dot b e}_\nu  \bar \sigma^{\dot c f}_\rho &
+\bar \sigma^{\dot a d}_\mu \bar \sigma^{\dot c e}_\nu  \bar \sigma^{\dot b f}_\rho 
+\bar \sigma^{\dot b d}_\mu \bar \sigma^{\dot a e}_\nu  \bar \sigma^{\dot c f}_\rho\\
 \bar \sigma^{\dot b d}_\mu \bar \sigma^{\dot c e}_\nu  \bar \sigma^{\dot a f}_\rho & 
+\bar \sigma^{\dot c d}_\mu \bar \sigma^{\dot a e}_\nu  \bar \sigma^{\dot b f}_\rho 
+\bar \sigma^{\dot c d}_\mu \bar \sigma^{\dot b e}_\nu  \bar \sigma^{\dot a f}_\rho
\Big]\times\\
 \times \sigma_{\delta f \dot c}   ~ (\bar \sigma_\lambda\sigma_\gamma)_{\dot b \dot a}  (\sigma_\alpha \bar\sigma_\beta)_{de} & ~  p^\mu p^\nu p^\rho q^\delta_1 q_2^\lambda q_2^\alpha q_1^\gamma q_1^\beta\\
= \frac{1}{2j+1}\frac{|c_\phi|^2}{48\Lambda^6} \Bigg\{ -\rm{Tr}\big[\bar\sigma_\mu\sigma_\alpha\bar\sigma_\beta\sigma_\nu\bar\sigma_\lambda\sigma_\gamma\big]\rm{Tr}\big[\bar\sigma_\rho\sigma_\delta\big] & -\rm{Tr}\big[\bar\sigma_\mu\sigma_\alpha\bar\sigma_\beta\sigma_\nu\bar\sigma_\delta\sigma_\rho\bar\sigma_\lambda\sigma_\gamma\big]\\
 + \rm{Tr}\big[\bar\sigma_\mu\sigma_\alpha\bar\sigma_\beta\sigma_\nu\bar\sigma_\gamma\sigma_\lambda\big]\rm{Tr}\big[\bar\sigma_\rho\sigma_\delta\big] &
 +\rm{Tr}\big[\bar\sigma_\beta\sigma_\alpha\bar\sigma_\mu\sigma_\lambda\bar\sigma_\gamma\sigma_\rho\bar\sigma_\delta\sigma_\nu\big]\\
 +\rm{Tr}\big[\bar\sigma_\rho\sigma_\delta\bar\sigma_\mu\sigma_\alpha\bar\sigma_\beta\sigma_\nu\bar\sigma_\gamma\sigma_\lambda\big] &
 -\rm{Tr}\big[\bar\sigma_\lambda\sigma_\gamma\bar\sigma_\rho\sigma_\delta\bar\sigma_\mu\sigma_\alpha\bar\sigma_\beta\sigma_\nu\big]\Bigg\}\times\\
 \times p^\mu p^\nu p^\rho q_1^\delta q_2^\lambda & q_2^\alpha q_1^\gamma q_1^\beta\\
 = \frac{1}{2j+1}\frac{|c_\phi|^2}{24\Lambda^6}m_{3/2}^2(m_{3/2}^2-m_H^2)^3. &
\eea
The decay rate now becomes:
\be
\Gamma(\psi_{3/2}   \to  H\bar\nu) 
    =  \frac{ 1 }{2j+1} \, \frac{|c_\phi|^2}{ 384 \pi}  \, \frac{m_{3/2}^7}{\Lambda^6} \, \left(1- \frac{M_H^2}{ m_{3/2}^2} \right)^4.
\ee

Finally, when computing the total decay width, one can use the fact that Lorentz symmetry demands that all spin states must have the same decay width. It is thus sufficient to compute the decay width for the highest spin state only and, the phase space integration will yield the spin averaged width. Using the highest spin state is convenient because the corresponding multispinor can be constructed from identical two-spinors, and thus, symmetrization is automatic.

%%%%%%%%%%%%%%%%%%%%%%%%%%%%%%%%%%%%%%  End %%%%%%%%%%%%%%%%%%%%%%%%%%%%%%%%%%%%%%%%%%%%%%%%

\bibliography{any_spin}

\end{document}